\documentclass[oldversion]{aa}

\usepackage{txfonts}
\usepackage{graphicx}
\usepackage{color}
\usepackage{natbib}
\bibpunct{(}{)}{;}{a}{}{,}

\newcommand{\ldisp}{\ensuremath{\sigma_\mathrm{line}\,}}
\newcommand{\mbh}{\ensuremath{M_{\mathrm{BH}}\,}}
\newcommand{\er}{\ensuremath{\lambda\,}}
\newcommand{\dd}{\mathrm{d}}

\begin{document}
\title{Low redshift AGN in the Hamburg/ESO Survey: II. The active black hole mass function and the distribution function of Eddington ratios
       \thanks{Based on observations collected at the European Southern Observatory, 
               Chile (Proposal 145.B-0009).} 
       }

\author{Andreas Schulze \and Lutz Wisotzki}
\institute{Astrophysikalisches Institut Potsdam, An der Sternwarte 16, 14482 Potsdam, Germany }
\offprints{A. Schulze, \email{aschulze@aip.de} }

\date{Received / Accepted }
\abstract{
We estimated black hole masses and Eddington ratios ($L/L_\mathrm{Edd}$) for a well defined sample of local ($z < 0.3$) broad line AGN from the Hamburg/ESO Survey (HES), based on the H$\beta$ line and standard recipes assuming virial equilibrium for the broad line region. The sample represents the low-redshift AGN population over a wide range of luminosities, from Seyfert 1 galaxies to luminous quasars. 

From the distribution of black hole masses we derived the active black hole mass function (BHMF) and the Eddington ratio distribution function (ERDF) in the local universe, exploiting the fact that the HES has a well-defined selection function. While the directly determined ERDF turns over around $L/L_\mathrm{Edd} \sim 0.1$, similar to what has been seen in previous analyses, we argue that this is an artefact of the sample selection. We employed a maximum likelihood approach to estimate the \emph{intrinsic} distribution functions of black hole masses and Eddington ratios simultaneously in an unbiased way, taking the sample selection function fully into account. The resulting ERDF is well described by a Schechter function, with evidence for a steady increase towards lower Eddington ratios, qualitatively similar to what has been found for type~2 AGN from the SDSS. 

Comparing our best-fit active BHMF with the mass function of inactive black holes we obtained an estimate of the fraction of active black holes, i.e.\ an estimate of the AGN duty cycle. The active fraction decreases strongly with increasing black hole mass. A comparison with the BHMF at higher redshifts also indicates that, at the high mass end, black holes are now in a less active stage than at earlier cosmic epochs. Our results support the notion of anti-hierarchical growth of black holes, and are consistent with a picture where the most massive black holes grew at early cosmic times, whereas at present mainly smaller mass black holes accrete at a significant rate.
}

\titlerunning{The local active black hole mass function}
\authorrunning{Schulze \& Wisotzki}
\maketitle

\section{Introduction}
The observed relations between the black hole mass and the properties of the spheroidal galaxy component imply a close connection between the growth of supermassive black holes (SMBH) and the evolution of their host galaxies. For local galaxies a strong correlation between the mass of the SMBH and the luminosity or mass of the bulge component \citep{Magorrian:1998,Marconi:2003,Haering:2004}, as well as with the stellar velocity dispersion \citep[e.g.][]{Ferrarese:2000,Gebhardt:2000, Tremaine:2002,Gultekin:2009} have been established. Semi-analytical and numerical simulations also show the importance of black hole activity and their corresponding SMBH feedback for galaxy evolution \citep[e.g.][]{DiMatteo:2005,Springel:2005,Croton:2006,Cattaneo:2006,Khalatyan:2007,Booth:2009}. It became clear that the central SMBH of a galaxy and especially its growth is an important ingredient for our understanding of galaxy formation and evolution. 

Therefore a complete census of the black hole population and its properties is required. Active black holes that will be observable as AGN are particularly important to study black hole growth.
A useful tool to study the AGN population is the luminosity function (AGNLF). The observed evolution of the AGNLF has been used to gain insight into the growth history of black holes  \citep[e.g.][]{Soltan:1982,Yu:2002,Marconi:2004,Merloni:2004,Shankar:2007}, and it became clear that most of the accretion occurs during bright QSO phases. But, using the AGNLF alone usually requires some additional assumptions, e.g. for the mean accretion rate, and thus is affected by uncertainties and degeneracies. 
Disentangling the AGNLF into the underlying distribution functions, namely the active black hole mass function (BHMF) and the distribution function of Eddington ratios (ERDF), is able to provide additional essential constraints on the growth of SMBHs. 

To understand the influence of black hole growth on galaxy evolution over cosmic time, first the properties of growing black holes in the local universe have to be well understood. Thus, it is important to derive black hole masses and accretion rates for large, well defined samples of AGN.
However, measuring black hole masses is much more  difficult than measuring luminosities. Black hole masses for large samples of AGN can not be measured directly, but only estimated, using locally established scaling relations. 

The best method to estimate \mbh for type~1 AGN is reverberation mapping of the broad line region \citep{Blandford:1982,Peterson:1993}. Assuming virial equilibrium black hole masses can be estimated by
$M_{\mathrm{BH}} = f R_{\mathrm{BLR}} \Delta V^2 / G$,
where $R_{\mathrm{BLR}}$ is the size of the broad line region (BLR), $\Delta V$ is the broad line width in km/s and $f$ is a scaling factor of order unity, which depends on the structure, kinematics and orientation of the BLR. Although the physics of the BLR is still not well understood and thus a source of uncertainty \citep[e.g.][]{Krolik:2001}, the validity of the virial assumption has been shown by the measurement of time lags and line widths for different broad lines in the same spectrum \citep{Peterson:2000,Onken:2002,Kollatschny:2003}. 

However, reverberation mapping requires extensive and meticulous observations and thus is not appropriate for large samples. Fortunately, a scaling relationship has been established between $R_{\mathrm{BLR}}$ and continuum luminosity of the AGN, $R_{\mathrm{BLR}} \varpropto L^\gamma$  \citep{Kaspi:2000,Kaspi:2005,Bentz:2006}. Thus it became possible to estimate \mbh from  single-epoch spectra for large samples, and has been used extensively in the previous years for large AGN samples \citep[e.g.][]{McLure:2004,Vestergaard:2004,Kollmeier:2006,Netzer:2007a,Shen:2007,Fine:2008,Gavignaud:2008, Trump:2009}.

For the measurement of the line width, different measures are commonly used, and it is unclear if one is superior to the others for estimating black hole masses. Most commonly used is the FWHM, but it has been suggested that the line dispersion $\sigma_{\mathrm{line}}$, i.e. the second central moment of the line profile, is a better measure of the line width \citep{Peterson:2004,Collin:2006}. The line dispersion is more sensitive to the wings of a line and less to the core, whereas for the FWHM the opposite is the case. An additional measure of line width used is the inter-percentile value \citep[IPV,][]{Fine:2008}.

The application of the virial method to large AGN samples allowed the estimation of the active BHMF \citep{McLure:2004,Shen:2007,Greene:2007a,Vestergaard:2008,Kelly:2008b,Vestergaard:2009}. 
A dataset that is perfectly suited to study especially low redshift AGN is provided by the Hamburg/ESO Survey (HES). In this paper we use a local AGN sample, drawn from the HES, to estimate their black hole masses and Eddington ratios, and construct the active black hole mass function as well as the distribution function of Eddington ratios. 

We first present our data and our treatment of the spectra. We estimate black hole masses and Eddington ratios from the spectra using the virial method. Next, we determine the active BHMF, taking care to account for sample selection effects, inducing a bias on the BHMF. Thereby, we not only constrain the local active BHMF but also put constraints on the intrinsic underlying distribution function of Eddington ratios. Finally, we discuss our results in the context of the local quiescent BHMF as well as that of other surveys.

Thoughout this paper we assume a Hubble constant of $H_0 = 70$ km s$^{-1}$ Mpc$^{-1}$ and cosmological density parameters $\Omega_\mathrm{m} = 0.3$ and $\Omega_\Lambda = 0.7$.

\section{The Sample}
The sample of low redshift AGN used in this study is drawn from the QSO catalogue of the Hamburg/ESO Survey \citep{Wisotzki:2000c}. For a more detailed description of the survey and the sample used, see our companion paper \citep[][hereafter Paper I]{Schulze:2009a}. Here we only give a short summary. 

The HES is a wide-angle, slitless spectroscopy survey, mainly for bright QSOs, carried out in the southern hemisphere, utilising photographic objective prism plates. The HES covers a formal area of $\sim$9500~deg$^2$ in the sky. After digitisation, slitless spectra in the range $3200\:$\AA{}~$\la\lambda\la 5200\:$\AA{} have been extracted from the plates. From these spectra type~1 AGN have been identified, based on their peculiar spectral energy distribution.
Follow-up spectroscopy has been carried out to confirm their QSO nature. The HES picks up quasars with $B \la 17.5$ at redshifts of up to $z\simeq 3.2$.

The Hamburg/ESO Survey yields a well-defined, flux-limited sample with a high degree of completeness. The survey covers a large area on the sky and the quasar candidate selection takes care to ensure that low redshift, low luminosity objects, i.e AGN with prominent host galaxies, are not systematically missed. As in Paper~I, we want to use this wide luminosity range at low redshift, which is unique for optical surveys, to study the low-redshift AGN population.

To construct such a local AGN sample we selected all AGN from the final HES catalogue (Wisotzki et al., in prep.) that belong to the `complete sample' and that are located at redshifts $z < 0.3$. The sample contains 329 type~1 AGN.
Spectra are available for most of the objects from the follow-up observations. For five objects, spectra were either missing in our database or they were of such poor quality that they were deemed not usable for our purposes. Thus our sample is $324/329 \approx 98.5$~\% complete in terms of spectroscopic coverage.

\section{Measurement of Emission Line Widths} \label{sec:width}
For the estimation of \mbh for our low redshift AGN sample, the broad line width of the H$\beta$, or alternatively the H$\alpha$, emission line has to be determined.
For the measurement of  the line widths of the H$\alpha$ and H$\beta$ emission lines we fitted the spectral region around these lines by analytic functions, i.e. by a multi-component Gaussian model plus continuum. Over this short wavelength range we approximated the underlying continuum as a straight line. The H$\alpha$ and H$\beta$ lines are fitted by one, two or, if required, by up to three Gaussians, based on visual inspection of the fits. Due to the limited resolution of the spectra the narrow line component could only be subtracted for a few lines, if a clear attribution of one fitting component to a narrow line component was possible. Thus a narrow component was only subtracted if clearly identified in the fit. Care has been taken to avoid contamination of the lines by contribution from the [\ion{O}{iii}] $\lambda\lambda$ 4959,5007 \AA\, lines and the \ion{Fe}{ii} emission to the H$\beta$ line, as well as from [\ion{N}{ii}] and [\ion{S}{ii}] to the H$\alpha$ line, by fitting them simultaneously with the Balmer lines. For details on the line fitting we refer to Paper I.

\begin{figure*}
\centering
\includegraphics[height=16cm,angle=-90]{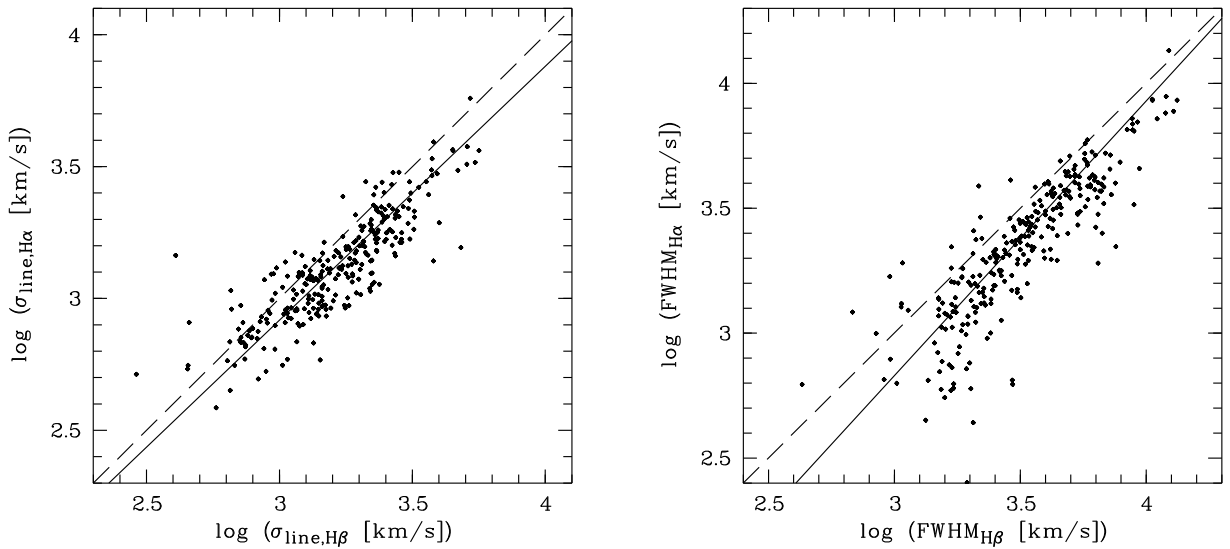}
\caption{Left panel: Correlation between $\sigma_{\mathrm{line}}(\mathrm{H}\beta)$ and $\sigma_{\mathrm{line}}(\mathrm{H}\alpha)$. Right panel: Correlation between FWHM(H$\beta$) and FWHM(H$\alpha$). The solid line shows the regression result using the FITEXY method and the dashed line is a one-to-one correspondence.}
\label{fig:reg}
\end{figure*}

We use two different line width measurements, the FWHM and the line dispersion for comparison, because there is at the moment no consensus which is the most appropriate for the estimation of black hole masses. Both can be easily derived from the fit. We then corrected the line widths for the finite resolution of the spectrograph.
We measured the continuum flux at 5100~\AA\, from the continuum fit to the H$\beta$ line region. We corrected the flux for Galactic extinction, using the dust maps of \citet{Schlegel:1998}, and the extinction law of \citet{Cardelli:1989} and computed the continuum luminosity $\lambda L_\lambda$(5100 \AA), hereafter $L_{5100}$.

For the estimation of errors we constructed artificial spectra for each object, using the fitted model and Gaussian random noise, corresponding to the measured S/N. We used 500 realizations for each spectrum. We fitted these artificial spectra, fitting the line and the continuum and measured the FWHM, the line dispersion and the line flux. The error was then simply taken as the dispersion between the various realizations. This method provides a formal error, taking into account fitting uncertainties caused by the noise. Thereby we assume that our multi-Gaussian fitting model provides a sufficiently precise model of the true line shape. Intrinsic deviations of the line shape from the model will increase the error. A remaining \ion{Fe}{ii} contribution at H$\beta$ might also increase the error.

For a subsample of 21 AGN also included in the SDSS Data Release 5 \citep[DR5; ][]{Adelman:2007}, we compared our results to the higher resolution SDSS spectra. We fitted the SDSS spectra in the same manner as the HES spectra. The correlation for $\sigma_{\mathrm{line}}$ is tight (we found a scatter of 0.07~dex for H$\beta$), whereas the scatter in the measurement of the FWHM is significantly larger (0.18~dex for H$\beta$). This is at least partially caused by the narrow component that can be disentangled better in the SDSS spectra. In contrast, $\sigma_{\mathrm{line}}$ is less susceptible to the narrow line contribution and thus provides a more precise width measurement for our sample. We also see evidence for an small underestimation of the line width compared to the SDSS spectrum, especially for narrower lines, with a mean deviation of 0.03~dex. This might be caused by the lower resolution of the HES spectra compared to the SDSS spectra and therefore the stronger influence of the resolution correction.

A comparison of the continuum luminosity, FWHM and black hole mass with the quasar sample of \citet{Shen:2007} is in general agreement with our values for the few objects in common.

The line dispersion is more sensitive to the wings of a line, thus to the subtraction of the contaminating lines, i.e. \ion{Fe}{ii} and [\ion{O}{iii}] for H$\beta$ and [\ion{N}{ii}] and [\ion{S}{ii}] for H$\alpha$. On the other hand, the FWHM is more susceptible to the line core, thus to a proper  subtraction of the narrow component \citep[see ][]{Denney:2008}. For our data the latter seems to exhibit the larger uncertainty. Together with the indication that $\sigma_{\mathrm{line}}$ is a preferable width estimate over the FWHM \citep{Peterson:2004,Collin:2006}, we decided to use $\sigma_{\mathrm{line}}$ to estimate black hole masses, and only give the results using the FWHM for comparison.

\subsection{Relations between H$\mathbf{\beta}$ and H$\mathbf{\alpha}$ Line Widths} 

We see a well-defined correlation between the line widths (both FWHM and \ldisp) of the H$\beta$ and H$\alpha$ emission lines, as shown in Fig.~\ref{fig:reg}. 
To quantify this relation, we applied a linear regression between H$\alpha$ and H$\beta$ in logarithmic units, using the FITEXY method \citep{Press:1992}, that accounts for errors in both coordinates.  We accounted for intrinsic scatter in the relation following \citet{Tremaine:2002} by increasing the uncertainties until a $\chi^2$ per degree of freedom of unity was obtained.

We found the following relations for the line widths:
\begin{equation}
\log \sigma_{\mathrm{H}\alpha} = 0.96  \log \sigma_{\mathrm{H}\beta} - 0.08  \ , \label{eq:rel_ldisp}
\end{equation} 
\begin{equation}
\log \mathrm{FWHM}_{\mathrm{H}\alpha} =  1.10  \log \mathrm{FWHM}_{\mathrm{H}\beta} - 0.17 \ . \label{eq:rel_fwhm}
\end{equation}
The rms scatter around the best fits are 0.11 dex for $\sigma_{\mathrm{line}}$ and 0.16 dex for the FWHM respectively. The relations between H$\alpha$ and H$\beta$ line properties using the FITEXY method are shown in Fig.~\ref{fig:reg}.

The relation obtained for the FWHM slightly deviates from the relations obtained by \citet{Greene:2005} and \citet{JShen:2008}, showing a stronger deviation from a one-to-one correlation. This might be due to the lower resolution of our data, thus the resolution correction has a stronger effect on the line width. This is supported by the slightly larger scatter for our relation. The scatter in the relation between the line dispersions is lower than between the FWHMs, again favouring \ldisp over FWHM for our data.

The H$\beta$ lines are on average broader than H$\alpha$ with $\langle \mathrm{FWHM}(\mathrm{H}\beta)/\mathrm{FWHM}(\mathrm{H}\alpha) \rangle =1.54$ and $\langle \sigma_{\mathrm{line}}(\mathrm{H}\beta)/\sigma_{\mathrm{line}}(\mathrm{H}\alpha) \rangle =1.29$ respectively. This is larger than found in other samples \citep[e.g.][]{Osterbrock:1982,Greene:2005} but in general agreement with the physical expectation of an increasing density or ionisation parameter of the BLR with decreasing radius.

\section{Results}

\subsection{Estimation of Black Hole Masses}

We estimated black hole masses for the AGN using the common scaling relationship. The sample of quasars analysed is well inside the ranges in redshift, with $z<0.3$, and in luminosity, with $10^{42}\leq L_{5100}\leq 10^{46}$ erg s$^{-1}$, over which the scaling relationship based on reverberation mapping has been established. So the estimated black hole masses do not suffer from an extrapolation of this relationship outside the range for which it is observationally tested.

For the scaling relationship between BLR size and continuum luminosity we use the values of \citet{Bentz:2008}:
\begin{equation}
\log R_{\mathrm{BLR}} = -21.3 + 0.519 \log ( L_{5100}) \ ,\label{eq:RBLR}
\end{equation}
with $L_{5100}$ given in erg s$^{-1}$ and $R_{\mathrm{BLR}}$ in light days. \\
The black hole mass is thus computed by
\begin{equation}
M_{\mathrm{BH}} =  6.7\cdot f \;\left( \frac{L_{5100}}{10^{44} \, \mathrm{erg s}^{-1}} \right)^{0.52} \left(  \frac{ \Delta V}{\mathrm{km/s}}\right)^2 M_{\sun} \, \label{eq:mbh2}
\end{equation}
where $f$ is the scale factor and $\Delta V$ the line width used, i.e. the FWHM or $\sigma_{\mathrm{line}}$. We computed black hole masses based on both width measurements. We prefer the black hole masses using the line dispersion and give the FWHM based black hole masses as a reference.  For the line dispersion we used a scale factor $f=3.85$, following \citet{Collin:2006}. This factor has been determined by setting the black hole masses, computed from the mean spectrum of the reverberation mapping sample and using the line dispersion, to the local $M_{\mathrm{BH}}-\sigma_\ast$ relation for quiescent galaxies, similar to the work of \citet{Onken:2004}. For the FWHM we used the common scale factor $f=3/4$ \citep[e.g.][]{Netzer:1990}  appropriate for a spherical BLR. 

\begin{figure}
\centering
\resizebox{\hsize}{!}{\includegraphics[angle=-90,clip]{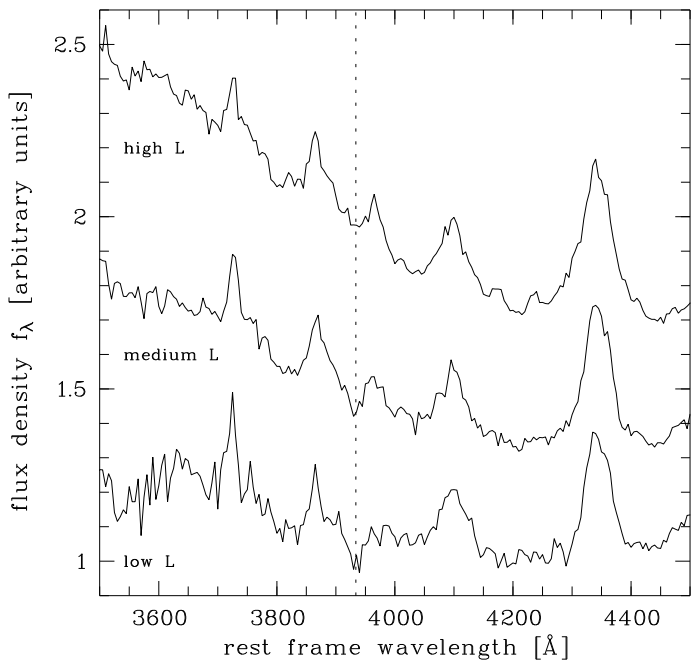}}
\caption{Median composite spectra for 3 luminosity bins, showing the \ion{Ca}{ii} K line region. The upper composite shows the high luminosity bin ($\log L_{5100}>44.5$), the middle composite is for the medium luminosity bin ($43.6<\log L_{5100}<44.5$) and the lower composite spectrum shows the low luminosity bin ($\log L_{5100}<43.6$). The \ion{Ca}{ii} K line at 3934~\AA{} is indicated as the dashed line.}
\label{fig:composites}
\end{figure}

\begin{figure}
\centering
\resizebox{\hsize}{!}{\includegraphics[angle=-90,clip]{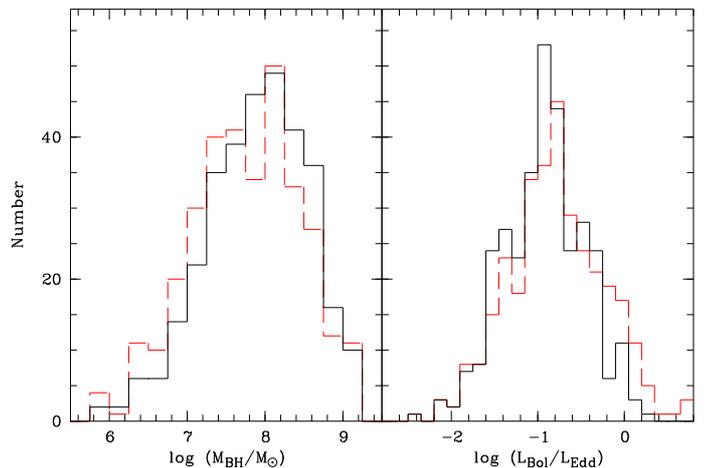}}
\caption{Left panel: Distribution of black hole masses. The black solid histogram shows the distribution of black hole masses estimated from $\sigma_{\mathrm{line}}$, the red dashed histogram shows $M_{\mathrm{BH}}$ estimated using the FWHM (and a constant scale factor of $f=0.75$). Right panel: Distribution of Eddington ratios. The histograms are the same as in the left panel.}
\label{fig:mbh}
\end{figure}

We have not corrected our continuum luminosities $L_{5100}$ for their host galaxy contribution. This might lead to an overestimation of \mbh for lower luminosity AGN, where the host contribution becomes significant. To disentangle the host contribution high resolution HST imaging is required, which is not available for our sample. However, the narrow slit used for the spectroscopy and the AGN selection technique already reduce the expected host contribution. Thus, the bias introduced by host galaxy contamination is expected to be small, but will lead to a systematic effect.

To estimate the degree of galaxy contribution to our AGN spectra we used the equivalent width (EW) of the \ion{Ca}{ii} K line at 3934~\AA, because this is the only prominent galaxy absorption feature not confused by AGN emission features within our spectral range. As this feature is only prominent in evolved stellar populations, the contribution from a very young stellar population might be neglected. However, low luminosity AGN hosts are known to have not particularly blue colours and generally do not show extremely young stellar populations, but rather indications for post starbursts \citep{VandenBerk:2006,Davies:2007}.

Since the mean signal-to-noise in our spectra is not sufficient to accurately measure the \ion{Ca}{ii} EW in individual spectra, we constructed composite spectra for three luminosity bins, depending on $L_{5100}$, shown in Fig.~\ref{fig:composites}. While no \ion{Ca}{ii} absorption is detected for the highest luminosity composite ($\log L_{5100}>44.5$), it is clearly present in the lower luminosity composites. We measure EWs of 0.8~\AA\, in the medium luminosity ($43.6<\log L_{5100}<44.5$) and of 2.0~\AA\, in the low luminosity ($\log L_{5100}<43.6$) median composite spectrum, respectively. 

To estimate the corresponding galaxy contribution, we used model spectra from single stellar population models with different ages and metallicities \citep{Bruzual:2003}. The low luminosity AGN, which will show the strongest host contribution, are preferentially spiral galaxies. We modeled them by stellar populations with ages between 900~Myr and 2.5~Gyr. 
We added various constant AGN contributions to the spectra and measured the resulting EWs of the AGN+galaxy spectra. To derive the galaxy contribution at 5100~\AA\, we assumed a flux ratio of the AGN of $f_{5100}/f_{3934}=0.64$. An EW of 2.0~\AA, as measured for our low luminosity subsample, corresponds to a host contribution to $L_{5100}$ of $35-40$~\%. This would reduce our black hole mass estimate by $0.10-0.12$~dex. The upper limit we can put on the host contribution is $\sim50$~\%, implying 0.16~dex for the \mbh estimation. The medium luminosity subsample shows an average host contribution of $15-20$~\%, corresponding to an overestimation of \mbh by $0.04-0.05$~dex. 

We used these estimates to apply average host corrections to the continuum luminosities and thus to the black hole masses. Although these corrections might be wrong in individual cases, for the sample as a whole the host contribution is thereby accounted for as good as possible for these data. We verified that our results are not qualitatively affected by applying or neglecting this correction. The quantitative change in the results is certainly very small.

The distributions of black hole masses using the FWHM and the line dispersion are shown in Fig.~\ref{fig:mbh}. The usage of the FWHM instead of $\sigma_{\mathrm{line}}$ slightly shifts the distribution to lower values, with $\langle \log M_{\mathrm{BH}} \rangle$ decreasing from 7.90 to 7.77, and also broadens the distribution, with the standard deviation changing from 0.65~dex for $\sigma_{\mathrm{line}}$ to 0.70~dex for FWHM.

In the following we will only refer to the black hole masses using $\sigma_{\mathrm{line}}$. This width estimate provides a more reliable width measurement for our data compared to the FWHM, as discussed in Section~\ref{sec:width}. We have verified that our results are fully consistent when using the FWHM instead.

\subsection{Eddington ratios}
To compute the Eddington ratio $\er = L_{\mathrm{bol}}/L_{\mathrm{Edd}}$, which can be understood as a normalised accretion rate, we estimated the bolometric luminosity from the optical continuum luminosity $L_{\mathrm{bol}} = f_L  L_{5100}$, applying a bolometric correction factor of $f_L=9$, as proposed by \citet{Kaspi:2000}. The mean bolometric correction factor is still somewhat uncertain, ranging from 7 \citep[e.g.][]{Netzer:2007a} to values around 10 \citep[e.g.][]{Richards:2006b,Marconi:2004}, and also seems to be dependent on luminosity \citep{Marconi:2004,Hopkins:2007}, black hole mass \citep{Kelly:2008c} or Eddington ratio \citep{Vasudevan:2009,Lusso:2009}. However, assuming a constant value $f_L$ at 5100 \AA\, is a good approximation. The value of $f_L=9$ is also in general agreement with the value obtained by integrating over the mean SED presented by \citet{Richards:2006b}.
The Eddington luminosity is given by $L_{\mathrm{Edd}} \cong 1.3 \cdot 10^{38} \left( M/M_\odot \right) \ \mathrm{erg \, s}^{-1}$.

The distribution of Eddington ratios, using the FWHM and $\sigma_{\mathrm{line}}$, are shown in the right panel of Fig.~\ref{fig:mbh}. The mean Eddington ratio of this sample is $\langle \log \er \rangle=-0.92$ with standard deviation of 0.46~dex using $\sigma_{\mathrm{line}}$, and $\langle \log \er \rangle=-0.79$ with 0.56~dex deviation for FWHM. 

This dispersion is higher than that found by other authors in higher redshift and higher luminosity samples \citep{Kollmeier:2006,Shen:2007}. Indeed, the shape of the observed distribution does depend on the underlying distribution function and the selection function of the survey. Thus the observed distribution of Eddington ratios is affected by the flux limitation of the survey and is not a quantity independent of the specific survey. The Eddington ratio distribution will change in mean and dispersion with luminosity \citep{Babic:2007,Hopkins:2008} due to this selection effect. Usually it will broaden with decreasing typical luminosity. 

This trend is also clearly visible in the sample of SDSS AGN presented by \citet{Shen:2007}. A redshift dependence is also indicated by their data. For their whole sample, covering the range $0.1\lesssim z \lesssim4.5$, they found a typical dispersion of $\sim 0.3$~dex, similar to the sample of \citet{Kollmeier:2006} that covers a similar redshift range and includes relatively high luminosity objects. Restricting the sample of \citet{Shen:2007} to $z\leq0.3$ gives a deviation of $0.43$~dex, similar to our results, but a lower mean Eddington ratio of -1.17 in logarithmic units. This trend is also present in deeper surveys that cover a wide redshift range. In the VVDS a value for the dispersion of $\sim 0.33$~dex has been found \citep{Gavignaud:2008}, while in the COSMOS survey a dispersion of $\sim 0.4$~dex has been observed \citep{Trump:2009}, in agreement with our low redshift result. We will discuss this issue further in Section~\ref{sec:intrinsic}.

\begin{figure*}
\centering
	\includegraphics[height=18cm,angle=-90]{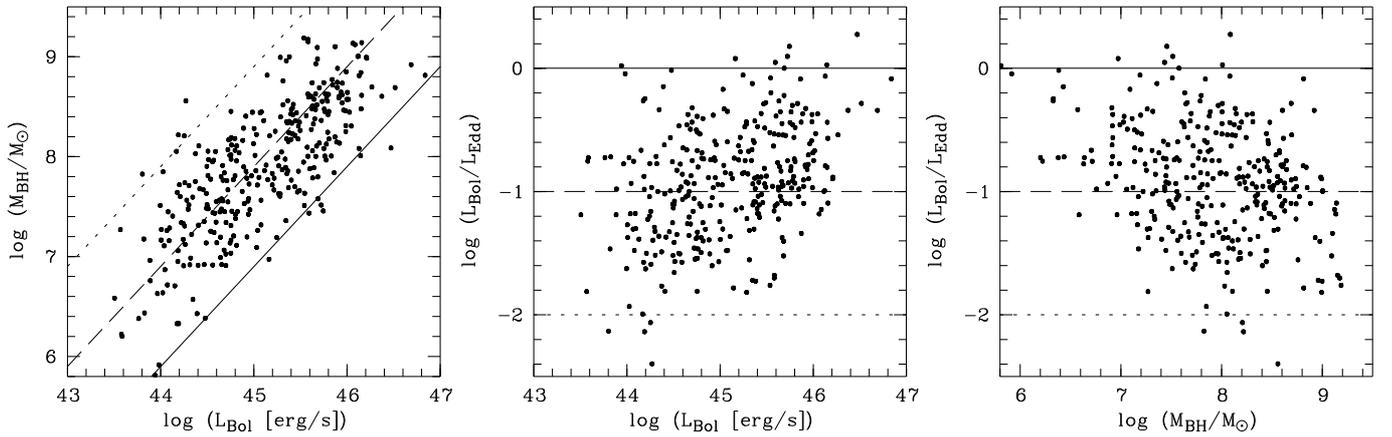}
\caption{Left panel: Black hole mass versus bolometric luminosity. Middle panel: Eddington ratio versus bolometric luminosity. Right panel: Eddington ratio versus black hole mass.
The three lines indicate Eddington ratios of 1 (solid), 0.1 (dashed) and 0.01 (dotted).}
\label{fig:Lbol}
\end{figure*}

\begin{figure}
\centering
\resizebox{\hsize}{!}{\includegraphics[height=16cm,angle=-90,clip]{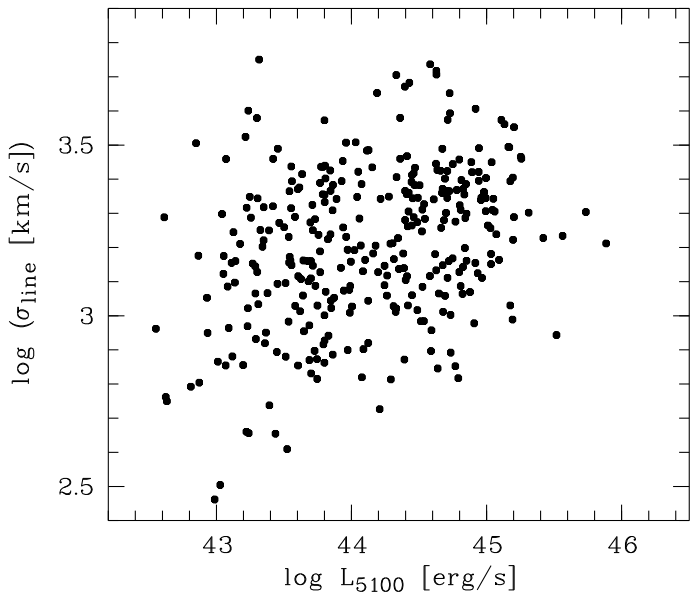}}
\caption{Distribution of the H$\beta$ line width ($\sigma_{\mathrm{line}}$) with continuum luminosity $L_{5100}$. There is only little correlation seen between line width and $L_{5100}$.}
\label{fig:L_width}
\end{figure}

In Fig.~\ref{fig:Lbol} we plot black hole mass, Eddington ratio and bolometric luminosity against each other. The first thing we have to be aware of when interpreting these plots are the implicit underlying correlations between these quantities. What we effectively always show is a combination of continuum luminosity $L_{5100}$ and line width \ldisp. Their underlying relation is shown in Fig.~\ref{fig:L_width}. There is only some week correlation present between $L_{5100}$ and line width.

Physically these plots can be understood from the shape of the underlying black hole mass function and distribution function of Eddington ratios  in combination with the selection function of the survey, as we will explicitly show in Section~\ref{sec:intrinsic}. The Eddington ratio \er spans the range $0.01< \er <1$, similar to other optical studies \citep{Woo:2002,Kollmeier:2006,Greene:2007a,Shen:2007}. At high values, observations have shown that the Eddington rate represents an approximate upper boundary to the Eddington ratio distribution, implying a steep decrease of the Eddington ratio distribution function toward Super-Eddington values. 
At low $\er$ the sample suffers from incompleteness due to the selection effects of the survey. This can explain the observed range of Eddington ratios and the rough correlation between \mbh and $L_\mathrm{bol}$, shown in the left panel of Fig.~\ref{fig:Lbol}.

No strong correlation is seen between \er and $L_\mathrm{bol}$ for this low redshift sample. There is a lack of objects in the lower right corner of the middle panel of Fig.~\ref{fig:Lbol}, thus a lack of objects with low-\er and high luminosity. These objects would have $\mbh > 2\cdot10^9 M_\odot$ and are rare objects due to the steep decrease of the black hole mass function at the high mass end (see Section~\ref{sec:bhmf} and \ref{sec:reconst}). Thus it is not surprising to see a lack of these objects in the sample. The same applies to the lack of objects seen in the upper right corner of the right panel of Fig.~\ref{fig:Lbol}. These would be objects with relative high \mbh and high Eddington ratio. This lack is also caused by the rarity of these objects, due to the steep decrease of the black hole mass function in combination with the decrease of the Eddington ratio distribution function toward the Eddington rate. Therefore, in the local universe massive black holes, accreting close to the Eddington limit, are exceedingly rare.

In the right panel of Fig.~\ref{fig:Lbol}, there is an absence of objects in the lower left corner, i.e. objects with low black hole mass and low Eddington ratio. These objects are victims of the survey selection.
They would have low luminosities and therefore only the closest would be detectable in a flux limited sample. An additional effect is that the AGN selection in the HES inevitably becomes incomplete at $M_{B_J}\gtrsim-19$, because the contribution of the host galaxy light even to the HES nuclear extraction scheme will become substantial, and the object will no longer be distinguished from a normal galaxy, due to a more galaxy like SED or due to a no longer detectable broad emission line. Therefore, no AGN with $M_{B_J}\gtrsim-18$ are detected in the survey, and the range $-18\gtrsim M_{B_J}\gtrsim-19$ is already seriously affected by this survey selection effect. Note that lines of equal luminosities in the right panel of Fig.~\ref{fig:Lbol} are diagonals from the upper left to the lower right. This selection effect explains the absence of observed objects in this region and results in the apparent anti-correlation between Eddington ratio and black hole mass, also seen in other samples \citep[e.g.][]{McLure:2004,Netzer:2007a}.

In Section~\ref{sec:intrinsic} we will explicitly show by Monte Carlo simulations how the observed distributions arise from an assumed underlying BHMF and Eddington ratio distribution function under consideration of the survey selection criteria.

\section{Black hole mass function and Eddington ratio distribution function} \label{sec:bhmf_erdf}

\subsection{The local active black hole mass function} \label{sec:bhmf}

The BHMF of quiescent galaxies in the local universe can be estimated, based on the relation between \mbh and bulge properties \citep[e.g.][]{Salucci:1999,Yu:2002,Shankar:2004,Marconi:2004}. Only a small fraction of local black holes are currently in an active state, accreting at a significant level and appearearing as an AGN. However, AGN do not accrete at a single value of $\er$, but rather show a wide distribution of  Eddington ratios \citep[e.g.][]{Heckman:2004,Yu:2005,Merloni:2008,Ho:2009,Kauffmann:2009}. Therefore it is not obvious what exactly to call an \textit{active} black hole. A pragmatic definition is to use a lower limit for the Eddington ratio. A natural choice for such a lower Eddington ratio for optical type~1 AGN samples would be $\er \simeq 0.01$, as this is approximately the observed lower value.

By this definition, our sample suffers from incompleteness at low black hole masses, because some low mass and low $\er$ AGN will be fainter than the flux-limit. The sample is not selected on black hole mass or Eddington ratio but on AGN flux. As already mentioned, the sample becomes incomplete at $M_{B_J}\gtrsim-19$. Thus, at low black hole mass only the AGN above this luminosity limit will be detected. This introduces a selection effect on the black hole mass distribution that needs to be taken into account for the determination of the BHMF. In the following, we will refer to this selection effect on the black hole mass and the Eddington ratio distribution as sample censorship, to distinguish it from more direct, for example redshift dependent, selection effects on the AGN luminosity distribution.

It is in principle possible to correct for this sample censorship by proper use of the respective selection function. If applying the usual selection function, which is a function of luminosity, and is appropriate for the determination of the luminosity function, to the determination of the black hole mass function, incompleteness is introduced because it has not properly accounted for active black holes below the flux limit \citep{Kelly:2008b}. Instead, the selection function has to be derived as a function of black hole mass and this selection function has to be applied to the construction of the BHMF. However, to do so would require knowledge of the, a priori unknown, Eddington ratio distribution function. Thus this approach is not feasible without additional assumptions. Nevertheless, it can be useful as a consistency check, as we will show in Section~\ref{sec:intrinsic_va}.
To avoid such additional assumptions, we used a different approach to determine the intrinsic underlying active BHMF from our data, taking into account the effect of sample censorship. These results are presented in  Section~\ref{sec:intrinsic}.

\begin{table*}
\caption{Fitting results for the local active black hole mass function, corrected for evolution but not for sample censorship.}
\label{tab:bhmf_fit}
\centering
\begin{tabular}{lrrrrrr}
\hline \hline \noalign{\smallskip}
Function & $\phi_\bullet^\ast$ in Mpc$^{-3}$ & $\log M_\ast$ & $\alpha$ & $\beta$ & $\chi^2$ & $\chi^2$/d.o.f. \\ 
\hline \noalign{\smallskip}
DPL                   & $2.86\times 10^{-6}$ & 7.86 & $-$0.74 & $-$3.11 &  7.29  & 0.81 \\ 
Schechter             & $2.73\times 10^{-6}$ & 8.06 & $-$0.84 & --      & 15.47  & 1.55 \\ 
mod. Schechter        & $4.96\times 10^{-6}$ & 6.97 & $-$0.25 &    0.51 & 13.64  & 1.52 \\ 
DPL (FWHM)            & $9.95\times 10^{-7}$ & 8.21 & $-$1.16 & $-$3.62 & 11.66  & 1.30 \\ 
Schechter (FWHM)      & $1.37\times 10^{-6}$ & 8.19 & $-$1.08 & --      & 16.40  & 1.64 \\ 
mod. Schechter (FWHM) & $5.15\times 10^{-6}$ & 6.81 & $-$0.59 &    0.32 & 15.33  & 1.70 \\ 
\noalign{\smallskip} \hline
\end{tabular}
\end{table*}

However, in this section we first determine the active BHMF, ignoring the effect of sample censorship on the data. We construct the BHMF using the usual selection function also used for the determination of the AGN luminosity function. However, it must be kept in mind that in this case we ignore active black holes with luminosities below the flux limit of the survey, even if their Eddington ratio is high enough to call it active by the above definition. Thus, this determined BHMF suffers from incompleteness at low mass caused by the sample censorship.
Nevertheless, this exercise is worthwhile, because it does not require any assumptions on the shape of the mass function or any information about the Eddington ratio distribution function. While the low mass end clearly will be affected by sample censorship, the high mass end is already well determined by this approach, providing important information on this mass range. Also, this uncorrected BHMF can be better compared with previous estimates on the BHMF that usually have not properly accounted for the sample censorship.

We constructed this active BHMF, not corrected for sample censorship, in an equivalent manner as  for the determination of a luminosity function (see Paper~I). We made use of the classical $1/V_\mathrm{max}$ estimator \citep{Schmidt:1968} to construct a binned BHMF. The differential BHMF (space density per log \mbh) is thus given by:
\begin{equation}
\Phi(M_\bullet) = \frac{1}{\Delta \log M_\bullet}\sum_{k} \frac{1}{V_{\mathrm{max}}^k}  \ ,
\end{equation}
where $V_{\mathrm{max}}$ is the maximal accessible volume in which the object, with given magnitude could have been found, given the flux-limit of the survey and the redshift bin used. The AGN sample used has been selected based on UV excess measured in the slitless spectra and no selection based on the presence of emission lines is applied. Thus, the $V_{\mathrm{max}}$ values used are equal to the ones used for the determination of the AGN luminosity function, presented in Paper~I. We lack usable spectra for 5 objects, and for an additional 7 objects we could not fit H$\beta$ due to poor quality of the spectra in this region and/or due to a low H$\beta$ contribution. Therefore we could not estimate \mbh for 12 objects. We took this into account in the survey selection function by multiplying the effective area by a factor of $317/329$. The exclusion of the 7 objects without proper H$\beta$ measurement may potentially bias our results. Therefore we estimated the H$\beta$ width from the H$\alpha$ measurement, using Equation~\ref{eq:rel_ldisp}, and then estimated \mbh for these 7 objects. Including these objects results in a consistent BHMF. Thus, the in- or exclusion of these objects makes no difference.

To derive the local ($z=0$) BHMF we corrected for evolution within our narrow redshift bin, $0<z<0.3$, as described in Paper~I. We applied a simple pure density evolution model within the redshift bin, i.e. $\rho(z) = (1+z)^{k_D}$ with $k_D=5$, thus adjusting our BHMF to redshift zero. This specific value ensures a result of the $V/V_\mathrm{max}$ test consistent with $\langle V/V_\mathrm{max}\rangle=0.5$, as would be expected in the case of no evolution.

\begin{figure}
\centering
\resizebox{\hsize}{!}{\includegraphics[angle=-90,clip]{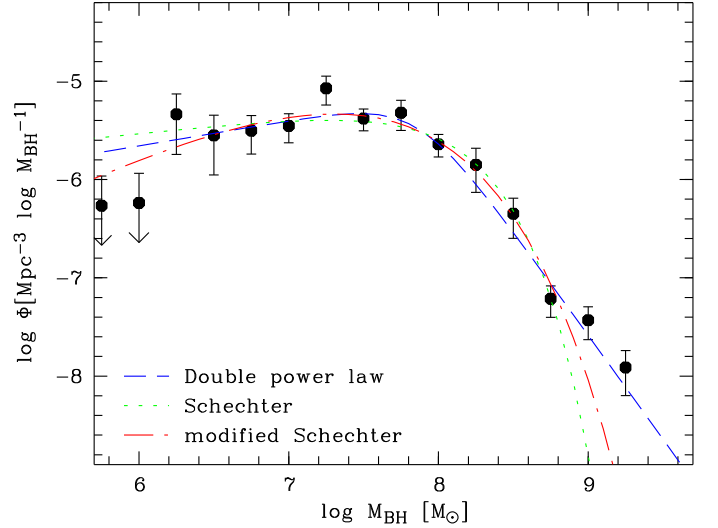}}
\caption{The differential active black hole mass function for $z=0$, not corrected for sample censorship. Filled black symbols show the BHMF using the line dispersion to estimate the black hole mass. The dashed line shows the double power law fit to the BHMF, the dotted line gives the Schechter function fit and the dashed dotted line represents the fit using a modified Schechter function.}
\label{fig:bhmf}
\end{figure}

The differential active BHMF of the HES is computed for bins of $\Delta \log \mbh =0.25$~dex in the range $10^{6}\leq \mbh \leq 10^{9.5}$. The resulting differential local BHMF, not corrected for sample censorship, is shown in Fig.~\ref{fig:bhmf}.

We used the following functional forms to fit the BHMF. A double power law, given by:
\begin{equation}
\phi(M_\bullet) = \frac{\phi^*/M_*}{(M_\bullet / M_*)^{-\alpha} + (M_\bullet / M_*)^{-\beta} } \ , \label{eq:dpl}
\end{equation}
where $M_*$ is a characteristic break mass, $\phi^*$ the normalisation and $\alpha$ and $\beta$ are the two slopes. A \citet{Schechter:1976} function, given by:
\begin{equation}
\phi(M_\bullet) = \frac{\phi^*}{M_*} \left( \frac{M_\bullet}{M_*} \right)^{\alpha} \exp \left( -\frac{M_\bullet}{M_*} \right)  \ . \label{eq:schecht}
\end{equation}
is also used.

We additionally used a functional form, motivated by the quiescent BHMF. The quiescent BHMF is given as a convolution of a Schechter function with a Gaussian and can be parameterised by the following function \citep[e.g.][]{Aller:2002,Shankar:2004}:
\begin{equation}
\phi(M_\bullet) = \frac{\phi^*}{M_*} \left( \frac{M_\bullet}{M_*} \right)^{\alpha} \exp \left( - \left[ \frac{M_\bullet}{M_*}\right]^{\beta}  \right)  \ .  \label{eq:modschecht}
\end{equation}
This basically corresponds to an ad-hoc modification of the Schechter function with an extra parameter $\beta$. A value $\beta>1$ corresponds to a decrease stronger than exponential and $\beta<1$ corresponds to a milder than exponential decrease. For $\beta=1$ this function turns into the usual Schechter function. In the following we refer to it as the modified Schechter function.

These BHMFs are connected to the expression in logarithmic units by $\Phi(M_\bullet)=(M_\bullet/\log_{10}e)\phi(M_\bullet)$.
The resulting fitting parameters of these three functions to our binned BHMF are listed in Table~\ref{tab:bhmf_fit}. All  give acceptable fits, while the Schechter function performs poorly at the highest black hole masses. However, the BHMF is less well constrained at high \mbh due to the small number of objects in these bins.

The shape of the BHMF is described by a steep decrease of the space density towards higher \mbh with $\beta \approx -3$ in the double power law, and a significant flattening at $\mbh \approx 10^{8} M_\odot$ toward lower \mbh. The high mass regime is not affected by the already mentioned sample censorship, while the low mass flattening is partially caused by the systematic underrepresentation of low $\er$ objects at low mass.

\begin{table}
\label{tab:bhmf}
\caption{Binned black hole mass function, not corrected for sample censorship. N gives the number of objects in each bin, $\log \phi$ and $\Delta \log \phi$ gives the space density per unit logarithmic black hole mass in solar masses and its 1$\sigma$ error respectively.}
\centering
\begin{tabular}{rrrrr} \hline\hline \noalign{\smallskip}
  & \multicolumn{2}{c}{\large{$\mathbf{\sigma}_\mathrm{line}$}}	& 
\multicolumn{2}{c}{\large{FWHM}} \\ \noalign{\smallskip}
$\log M_{\mathrm{BH}}$ & $N$ & $\log \Phi_M$  & $N$ & $\log \Phi_M$ \\ 
\hline \noalign{\smallskip}
5.75 & 1 & $-$6.27$^{+0.31}_{-\infty}$ & 4 & $-$5.75$^{+0.20}_{-0.36}$ \\ \noalign{\smallskip} 
6.00 & 1 & $-$6.24$^{+0.30}_{-\infty}$ & 0 &  \\ \noalign{\smallskip} 
6.25 & 4 & $-$5.34$^{+0.21}_{-0.40}$ & 6 & $-$5.33$^{+0.20}_{-0.41}$ \\ \noalign{\smallskip} 
6.50 & 5 & $-$5.55$^{+0.20}_{-0.40}$ & 8 & $-$5.49$^{+0.19}_{-0.33}$ \\ \noalign{\smallskip} 
6.75 & 6 & $-$5.50$^{+0.15}_{-0.24}$ & 11 & $-$5.29$^{+0.15}_{-0.22}$ \\ \noalign{\smallskip} 
7.00 & 20 & $-$5.45$^{+0.12}_{-0.18}$ & 28 & $-$5.34$^{+0.11}_{-0.15}$ \\ \noalign{\smallskip} 
7.25 & 27 & $-$5.07$^{+0.12}_{-0.17}$ & 35 & $-$5.15$^{+0.12}_{-0.16}$ \\ \noalign{\smallskip} 
7.50 & 38 & $-$5.38$^{+0.10}_{-0.12}$ & 42 & $-$5.65$^{+0.09}_{-0.12}$ \\ \noalign{\smallskip} 
7.75 & 47 & $-$5.32$^{+0.13}_{-0.18}$ & 33 & $-$5.45$^{+0.15}_{-0.22}$ \\ \noalign{\smallskip} 
8.00 & 42 & $-$5.64$^{+0.10}_{-0.13}$ & 45 & $-$5.63$^{+0.13}_{-0.19}$ \\ \noalign{\smallskip} 
8.25 & 42 & $-$5.85$^{+0.17}_{-0.28}$ & 36 & $-$5.78$^{+0.15}_{-0.24}$ \\ \noalign{\smallskip} 
8.50 & 45 & $-$6.35$^{+0.16}_{-0.25}$ & 32 & $-$6.37$^{+0.16}_{-0.27}$ \\ \noalign{\smallskip} 
8.75 & 21 & $-$7.21$^{+0.13}_{-0.19}$ & 23 & $-$7.20$^{+0.13}_{-0.18}$ \\ \noalign{\smallskip} 
9.00 & 13 & $-$7.43$^{+0.13}_{-0.20}$ & 8 & $-$7.62$^{+0.17}_{-0.29}$ \\ \noalign{\smallskip} 
9.25 & 5 & $-$7.91$^{+0.17}_{-0.29}$ & 6 & $-$7.83$^{+0.16}_{-0.25}$ \\ \noalign{\smallskip} 
\hline
\end{tabular} 
\end{table}

\subsection{The local Eddington ratio distribution function} \label{sec:erdf}

\begin{figure}
\centering
\resizebox{\hsize}{!}{\includegraphics[angle=-90,clip]{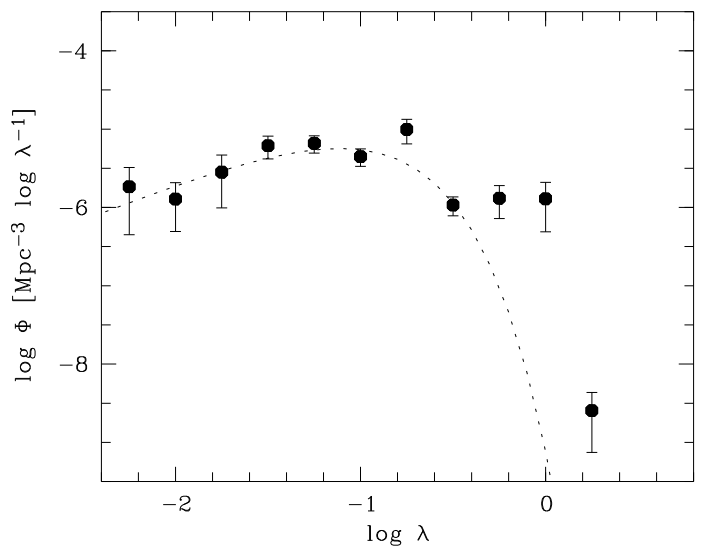}}
\caption{The differential Eddington ratio distribution function for $z=0$, not corrected for sample censorship. The dashed line shows the best Schechter function fit.}
\label{fig:arf}
\end{figure}

Given the estimates of the Eddington ratio $\er$ for our sample, we can analogously determine the local Eddington ratio distribution function (ERDF) for the HES, equivalent to the BHMF. This determination also does not take into account the effect of sample censorship.
We computed the local ERDF in bins of $\Delta \log \er =0.25$~dex in the range $-2.25\leq \log \er \leq 0.25$. The resulting differential local ERDF is shown in Fig.~\ref{fig:arf}.

The uncorrected AGN space density declines at high as well as at low $\er$, showing a peak around $\log \er \approx -1.0$. We fitted the ERDF by a Schechter function, neglecting the lowest $\er$ point. The resulting best fit values are $\phi_\er^\ast = 6.66 \times 10^{-6}$~Mpc$^{-3}$,  $\log \er_\ast = -1.01$ and $\alpha_\er=-0.05$ with a value of $\chi^2$ per degree of freedom of $1.9$.

However, also this ERDF is strongly affected by sample censorship. While at the highest Eddington ratios ($\log \er > -1$) the majority of AGN will be detected by the survey, at low Eddington ratio ($\log \er < -1$) a significant number of objects will have a too low luminosity to be detected. Therefore the space density at low $\er$ will be underestimated by the derived ERDF. In the next section we will reconstruct the intrinsic underlying ERDF as well as the intrinsic BHMF.

\section{Reconstruction of the intrinsic BHMF and ERDF} \label{sec:intrinsic}
\subsection{Method}   \label{sec:method}
As already noted, the BHMF presented so far is basically luminosity limited and thus incomplete at low mass in terms of an accretion rate limited active BHMF. We now want to constrain the intrinsic active BHMF by our observations, correcting for this sample censorship. We use $\log \er = -2$ as the lower limit for the Eddington ratio; for $\log \er = -2$ we call black holes 'active'.

The selection function of the survey is a function of luminosity, and thus of the product of \mbh and $\er$. Therefore, the reconstruction of the active BHMF also requires the knowledge of the ERDF.
Both distribution functions cannot be determined independently from each other. In Section~\ref{sec:intrinsic_va}, as a consistency test, we will determine the active BHMF assuming a specific ERDF. But without such an assumption both distribution functions have to be determined at the same time. This is the approach we will follow in this section. 

Knowing both distribution functions, the AGN luminosity function is directly given as a convolution of the two:
\begin{equation}
\Phi( L)= \int_{M_{\bullet,\mathrm{min}}}^\infty P_\er(\er)\Phi_\bullet(M_\bullet)\, \dd\log M_\bullet \ , \label{eq:lf}
\end{equation} 
where we adopt $\log M_{\bullet,\mathrm{min}}=6$. With $P_\lambda$ we define the normalised ERDF, thus:
\begin{equation}
P_\er(\er)=\frac{\Phi_\er(\er)}{  \int \Phi_\er(\er)\, \dd\log \er }\ . \label{eq:P_er}
\end{equation} 

We determined the BHMF and ERDF together, performing a maximum likelihood fit to the data \citep[e.g.][]{Marshall:1983}. We consider the joint Poisson probability distribution of black hole mass and Eddington ratio. We minimise the function $S=-2\ln \mathcal{L}$, with $\mathcal{L}$ being the likelihood of finding the observed data, given the respective model. Thus, $S$ is given by:
\begin{equation}
S = - 2 \sum_{i=1}^N \ln p(M_{\bullet,i},\er_i) + 2 \iint p(M_\bullet,\er) \dd\log \er \dd\log M_\bullet \ . \label{eq:maxlike}
\end{equation} 
The sum is over the observed objects and the integral is equal to the expected number of objects, given the assumed BHMF and ERDF. The probability distribution $p(M_\bullet,\er)\dd\log \er \dd\log M_\bullet$ gives the probability of finding an AGN with black hole mass between $\log M_\bullet$ and $\log M_\bullet+\dd\log M_\bullet$ and Eddington ratio between $\log \er$ and $\log \er+\dd\log \er$ in an observed sample. The total number of objects $N$ is then given by integration of $p(M_\bullet,\er)$ over $M_\bullet$ and $\er$.

We will now briefly motivate the used probability distribution $p(M_\bullet,\er)$ for our sample.
The observed number of objects in a sample is given by:
\begin{equation}
N = \iint \Omega_{\mathrm{eff}}(m,z) \, \Phi_L(\log L,z)\, \frac{\dd V}{\dd z}\,  \dd\log L\, \dd z\ ,
\end{equation}
where $\Phi_L(\log L)$ is the AGN luminosity function and $\Omega_{\mathrm{eff}}$ is the effective survey area as a function of apparent magnitude and redshift, thus the selection function for our flux-limited survey. This can be understood as a selection function depending on $z$, $M_\bullet$ and $\er$, thus $\Omega_{\mathrm{eff}}(m,z)=\Omega_{\mathrm{eff}}(L,z)=\Omega_{\mathrm{eff}}(M_\bullet,\er,z)$. 
For the Hamburg/ESO Survey the selection function within our covered redshift range is almost independent of redshift. Thus, as discussed in Paper~I, we can marginalise over redshift. For details on the selection function of the HES see \citet{Wisotzki:2000c}.

Apart from the flux limit, our sample is incomplete at the lowest luminosities $M_{B_J}>-19$. For low luminosity AGN the host galaxy contribution becomes an important factor and the objects might no longer be classified as an AGN, due to the SED being dominated by starlight. As shown in Paper~I, the sample is highly complete brighter than $M_{B_J}\approx-19$. Thus, we adopted a luminosity limit of $M_{B_J}<-19$ in the selection function, $\Omega_{\mathrm{eff}}$. We also restricted the observed sample to this lower luminosity for the comparison of the sample properties.

The AGN luminosity function  $\Phi_L(\log L)$ is related to the BHMF and the ERDF via Equation~\ref{eq:lf}.
For the redshift evolution, we assumed the simple pure density evolution model of Section~\ref{sec:bhmf}. In this case the black hole mass function is separable into a function of $M_\bullet$ and a function of $z$, $\Phi(M_\bullet,z)=\Phi(M_\bullet)\rho(z)=\Phi(M_\bullet)(1+z)^{k_D}$, with $k_D=5$. 

The expected number of objects for a given survey and an assumed BHMF and ERDF is then given by:
\begin{equation}
N = \iiint \Omega_{\mathrm{eff}} \, P_\er(\er)\Phi_\bullet(M_\bullet) \,(1+z)^{k_D} \frac{\dd V}{\dd z} \dd\log \er \dd\log M_\bullet \dd z \ . \label{eq:nobs}
\end{equation}
Thus, the bivariate probability distribution of black hole mass and Eddington ratio is given by:
\begin{equation}
p(M_\bullet,\er) = \int \Omega_{\mathrm{eff}} \, P_\er(\er)\Phi_\bullet(M_\bullet) \,(1+z)^{k_D} \frac{\dd V}{\dd z} \dd z \ . \label{eq:pbher}
\end{equation}

Given this bivariate distribution for an assumed BHMF and ERDF, we minimise the likelihood function $S$ (Equation~\ref{eq:maxlike}) using a downhill simplex algorithm \citep{Nelder:1965}. As a lower limit for the fitting we employed a black hole mass of $M_\mathrm{min}=10^6 M_\odot$ and an Eddington ratio of $\er_\mathrm{min}=10{-2}$. The HES sample was restricted to these limits accordingly.

For the BHMF we assumed three different models. Firstly we used a double power law with the high mass slope fixed to the value $\beta_\mathrm{BH} = -3.01$, determined from the uncorrected BHMF in Section~\ref{sec:bhmf}. This lowers the required number of free parameters and is justified, because the high mass region in the uncorrected BHMF is only weakly affected by incompleteness. Secondly we also used a double power law, but leaving the high mass slope as a free parameter, to be determined in the fit. As third model we used the function given by Equation~\ref{eq:modschecht}, thus a modified Schechter function. The starting values for the minimisation algorithm are taken from the fit to the uncorrected BHMF.

We decided to model the ERDF by a Schechter function, corresponding to an exponential cutoff close to the Eddington limit and a wide power law-like distribution at low Eddington ratio. This parameterisation differs from the often assumed log-normal distribution. However, a log-normal distribution is only motivated by the \emph{observed} distribution, not accounting for any selection effects. Also, a log-normal distribution enforces a maximum and a turnover at low $\er$. A Schechter function is more flexible, allowing for a turnover at low values, but not enforcing it. In particular, it allows an increase of the space density at low $\er$. This shape would be consistent with observations of type~2 AGN \citep{Yu:2005, Hopkins:2008, Kauffmann:2009}, with estimates for the total AGN population \citep{Merloni:2008} as well as with model expectations of AGN lightcurves from self-regulated black hole growth \citep{Yu:2008, Hopkins:2008}. 
Aside from the Schechter function parameterisation of the ERDF, we additionally tested a log-normal ERDF as functional form. Together with the Schechter function it covers a wide range of possible parameterisations for the ERDF.

From our data we are not able to constrain a dependence of the ERDF on \mbh, so we assumed the ERDF to be independent of \mbh, already implicitly assumed in Equation~\ref{eq:lf}. The normalisation of the ERDF is fixed by the condition that the BHMF and ERDF have to predict the same space density of AGN. This leaves two free parameters for the ERDF, the break $\er_\ast$ and the low-$\er$ slope $\alpha_\er$ for the Schechter function, or the mean $\er_\ast$ and the width $\sigma_\er$ for the log-normal distribution. However, these two parameters in both cases are not independent from each other, because the data by construction needs to be consistent with the observed luminosity function (LF). Thus, for a given BHMF and a fixed value for $\er_\ast$, $\alpha_\er$ is given by the condition that the LF derived from the BHMF and ERDF by Equation~\ref{eq:lf} has to be consistent with the observed LF.
Our approach automatically ensures the consistency of the BHMF and the ERDF with the observed LF.

To assess the goodness of fit for the individual models we used two different methods. This is required because the maximum likelihood method does not provide its own assessment of the goodness of fit. First, we used a two-dimensional K-S test \citep{Fasano:1987} on the unbinned data. Second, we employed a $\chi^2$ test, binning the data in bins of $0.5$~dex in \mbh and $\er$ respectively. The results are given with the best fit parameters in Table~\ref{tab:res}.

\begin{table*}
\caption{Fitting results for the active black hole mass function and the Eddington ratio distribution function. The first column indicates the function used for the BHMF. 'DPL' is for a double power law, with $\beta$ indicating the fixing of the high mass slope and 'mS' is for a modified Schechter function. The second column indicates the ERDF. 'S' is for a Schechter function and 'ln' stands for a log-normal distribution.}
\label{tab:res}
\centering
\begin{tabular}{llrrrrrrrrrrr}
\hline \hline \noalign{\smallskip}
\multicolumn{12}{c}{}& \multicolumn{1}{c}{$\rho_\mathrm{act}$} \\
BHMF & ERDF & $\phi_\bullet^\ast$ [Mpc$^{-3}$] & $\log M_\ast$ & $\alpha_\mathrm{BH}$ & $\beta_\mathrm{BH}$ & $\log \er_\ast$ & $\alpha_\er$/$\sigma_\er$ & $D_\mathrm{KS}$ & $p_\mathrm{KS}$ & $\chi^2$/d.o.f. & $p_{\chi^2}$ & \multicolumn{1}{c}{[$M_\odot \mathrm{Mpc}^{-3}$]} \\ 
\noalign{\smallskip} \hline \noalign{\smallskip}
DPL($\beta$) & S & $2.97\times 10^{-6}$ & 7.97 & $-$2.11 & $-$3.11 & $-$0.57 & $-$1.90 & 0.100 & 2.8e$-$2 & 61.3/25 & 4.2e$-$5 & 1621 \\  \noalign{\smallskip}
DPL & S & $2.86\times 10^{-6}$          & 8.01 & $-$2.10 & $-$3.21 & $-$0.56 & $-$1.94 & 0.101 & 2.6e$-$2 & 63.4/25 & 2.1e$-$5 & 1687 \\ \noalign{\smallskip} 
mS & S & $2.75\times 10^{-6}$           & 8.11 & $-$2.11 &    0.50 & $-$0.55 & $-$1.95 & 0.094 & 4.8e$-$2 & 56.8/25 & 1.8e$-$4 & 1767 \\ \noalign{\smallskip} 
mS & ln & $2.36\times 10^{-6}$          & 8.07 & $-$2.12 &    0.48 & $-$1.83 &    0.49 & 0.081 & 1.2e$-$1 & 50.8/25 & 1.1e$-$3 & 1388 \\ \noalign{\smallskip} 
\hline
\end{tabular}
\end{table*}

\begin{figure*}
\centering 
        \includegraphics[width=9cm]{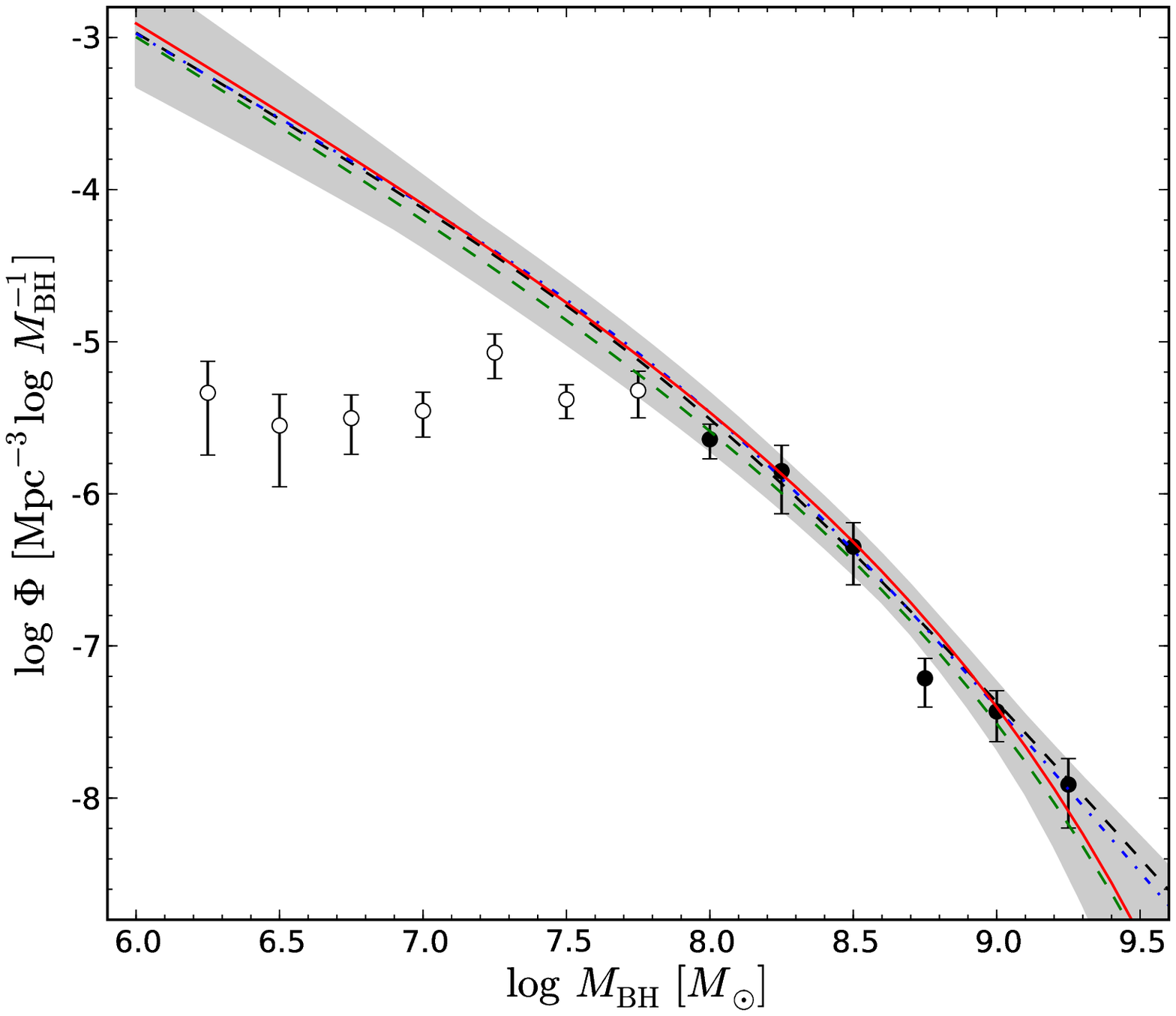}
        \includegraphics[width=9cm]{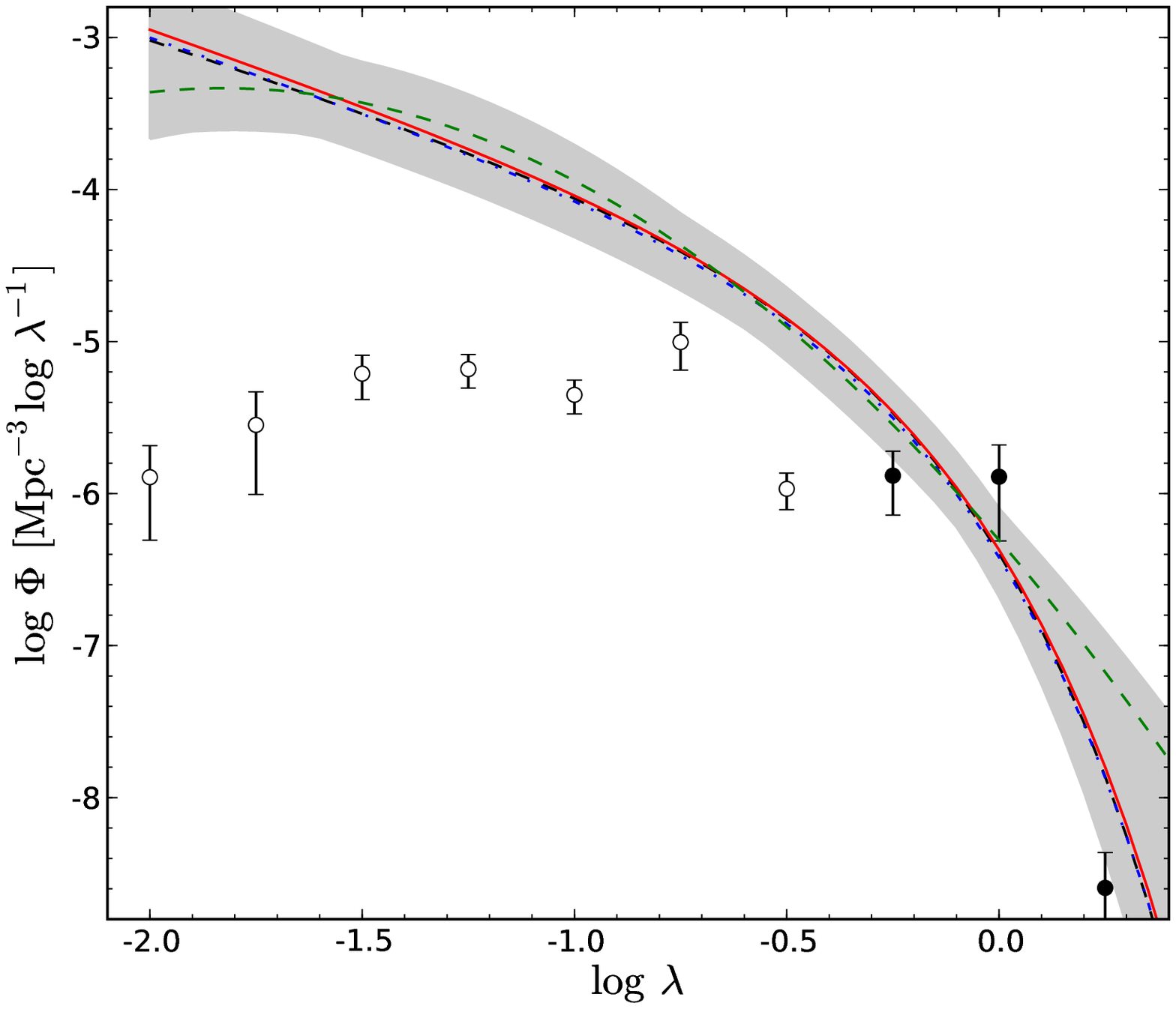}
\caption{Results for the reconstructed BHMF and ERDF. The left panel gives the BHMF and the right panel the ERDF respectively. The black points show the binned uncorrected distribution function, with filled circles representing bins that do not suffer significantly from sample censorship and open circles represent bins, biased by sample censorship. They are shown for comparison with the reconstructed BHMF and ERDF.
The black dashed line shows a double power law BHMF with fixed high mass slope $\beta=-3.11$ and Schechter ERDF, the blue dashed dotted line  is for a free double power law BHMF and Schechter ERDF, the red solid line represents a modified Schechter BHMF and Schechter ERDF and the green dashed line is for a modified Schechter BHMF and log-normal ERDF. The gray areas show the $1\sigma$ confidence regions of both distribution functions, taking into account all 4 parametric models.}
\label{fig:trueBhmf}
\end{figure*}

\subsection{Results} \label{sec:reconst}
The first model consists of a double power law BHMF, with the high mass slope fixed to $\beta=-3.11$, and a Schechter function ERDF. The best fit distribution functions are shown as black dashed line in Fig.~\ref{fig:trueBhmf} and their fit parameters are given in Table~\ref{tab:res}. The BHMF shows a steep high mass slope with $\alpha_\mathrm{BH}\approx-2$ and the break is consistent with the uncorrected BHMF. The ERDF is increasing towards low $\er$ down to the applied limit of $\er=0.01$. 

This function provides a good fit to the high mass end of the uncorrected BHMF, which is only little affected by sample censorship. At the low mass end the uncorrected BHMF strongly underpredicts the active black hole space density, compared to the reconstructed underlying active BHMF. This also holds true for all other applied functional forms for the BHMF and the ERDF. The same also applies to the uncorrected ERDF. The uncorrected ERDF is strongly biased and underestimates the BH space density. The best fit to the \emph{uncorrected} BHMF and to the ERDF is clearly rejected by the maximum likelihood approach with high confidence. They are not able to produce the observed distributions of \mbh and $\er$ and are not consistent with the AGN LF. 
This clearly shows that the usual approach used to construct an uncorrected BHMF and ERDF is strongly biased.

We briefly want to illustrate how the maximum likelihood approach  is able to reject certain models for the BHMF and ERDF and favour others. To compute the expected distributions within a grid of free parameters, we restricted the number of parameters to two. We fixed the break and normalisation of the BHMF. Thus, with the high mass slope already fixed, the only free parameter for the BHMF is the low mass slope $\alpha_\mathrm{BH}$. For the ERDF there are two free parameters, the break and the low-$\er$ slope of the Schechter function. However, one of these is fixed by the constraint to recover the observed AGN LF. We took $\alpha_\er$ as a free parameter and determined the break by a $\chi^2$ minimisation of the LF computed via Equation~\ref{eq:lf} to the observed LF. The normalised \textit{observed} distribution of $\log M_\bullet$ and $\log \er$ are given by:
\begin{equation}
p(\log M_\bullet) = \frac{1}{N}\int p(M_\bullet,\er) \dd\log \er  \label{eq:plogm}
\end{equation}
\begin{equation}
p(\log \er) = \frac{1}{N}\int p(M_\bullet,\er) \dd\log M_\bullet  \ . \label{eq:ploger}
\end{equation}

\begin{figure*}
\centering
\newcommand{\wi}{5.2cm}
\setlength{\unitlength}{1mm}
\begin{picture}(190,100)
\put(1,17){\rotatebox[origin=c]{90}{$\alpha_\lambda=-2.3$}}
\put(1,47){\rotatebox[origin=c]{90}{$\alpha_\lambda=-1.9$}}
\put(1,77){\rotatebox[origin=c]{90}{$\alpha_\lambda=-1.2$}}

\put(27,99){$\alpha_\mathrm{BH}=-2.8$}
\put(85,99){$\alpha_\mathrm{BH}=-2.1$}
\put(143,99){$\alpha_\mathrm{BH}=-1.3$}

\put(5,64){\includegraphics[width=\wi]{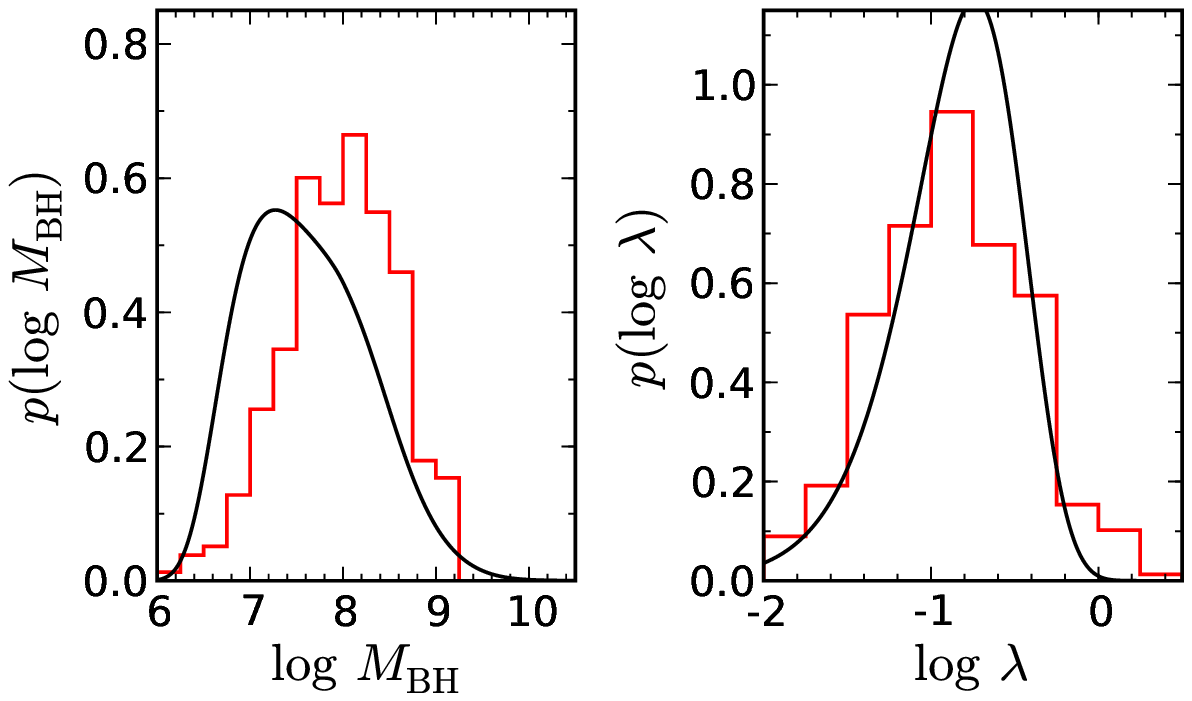}}
\put(63,64){\includegraphics[width=\wi]{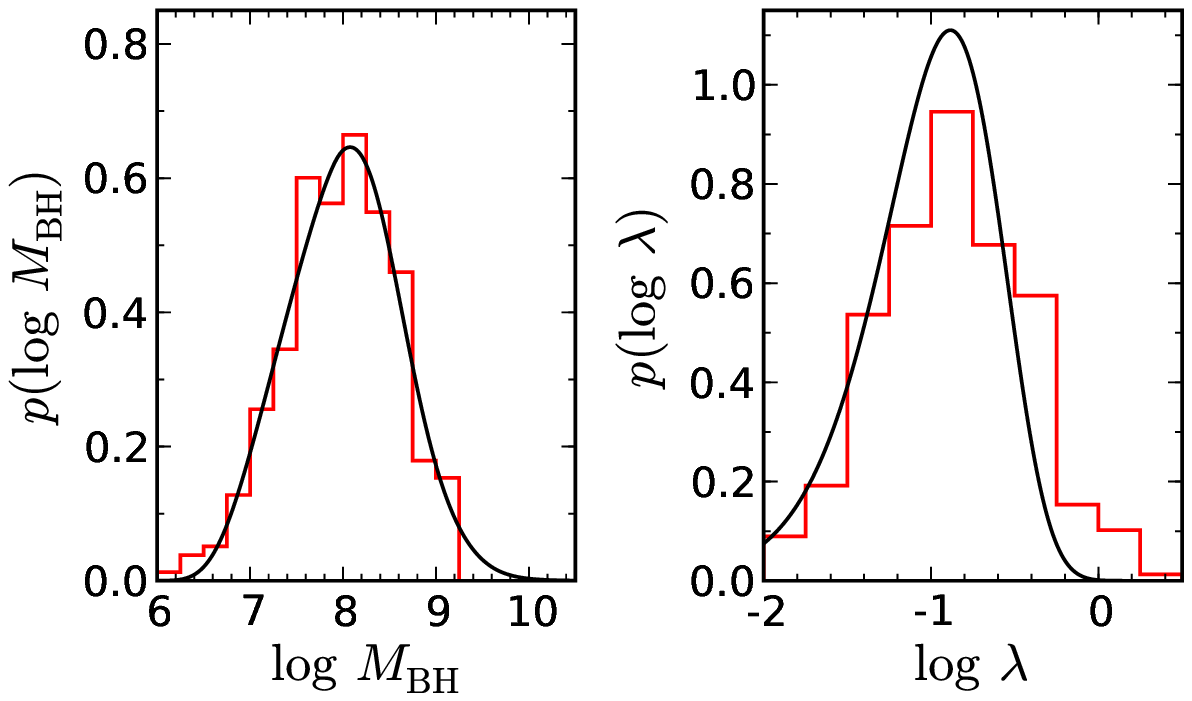}}
\put(121,64){\includegraphics[width=\wi]{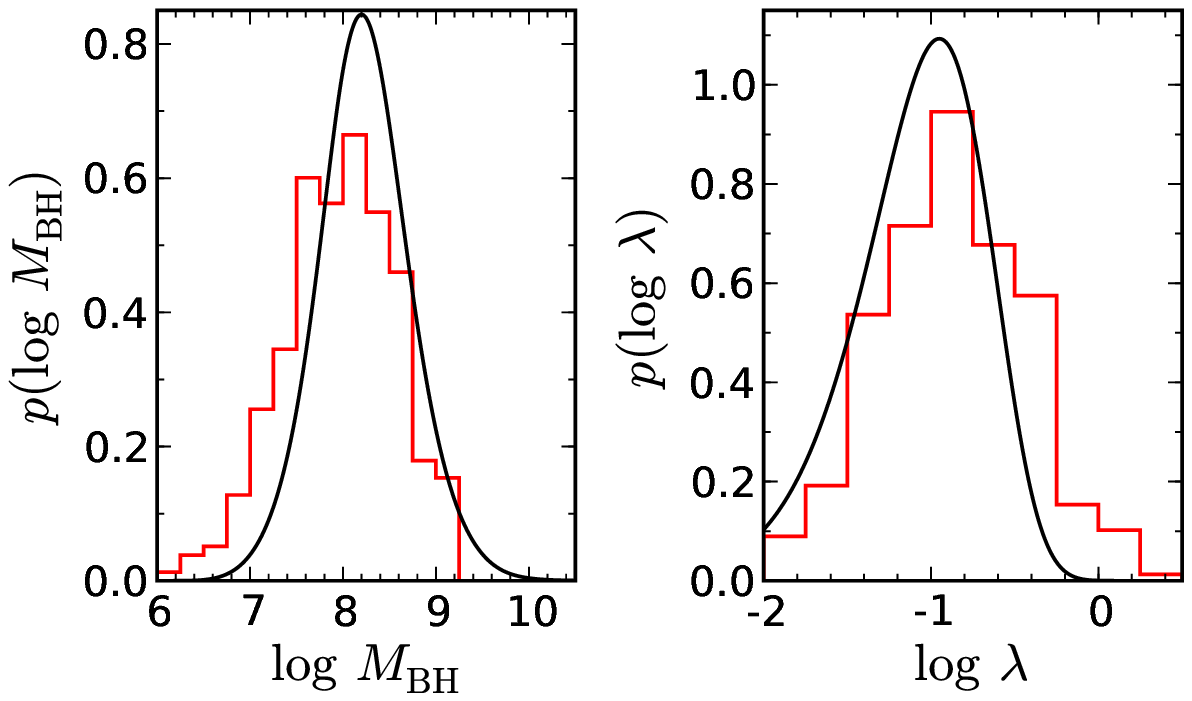}}

\put(5,32){\includegraphics[width=\wi]{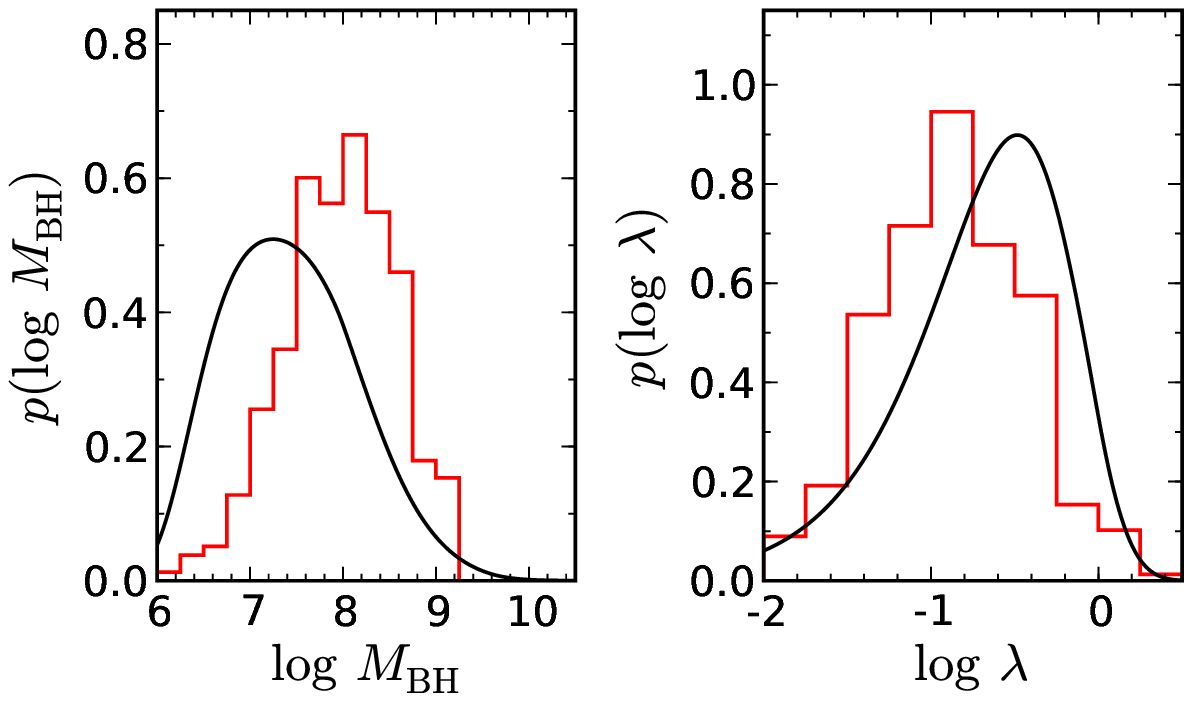}}
\put(63,32){\includegraphics[width=\wi]{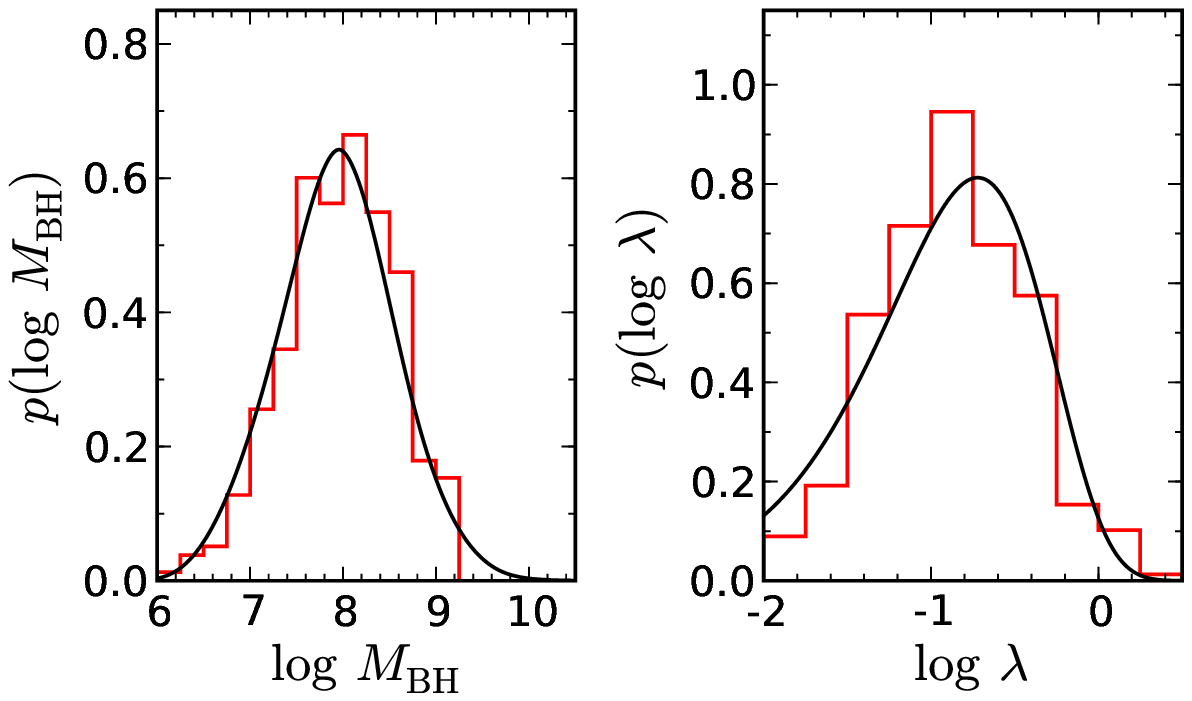}}
\put(121,32){\includegraphics[width=\wi]{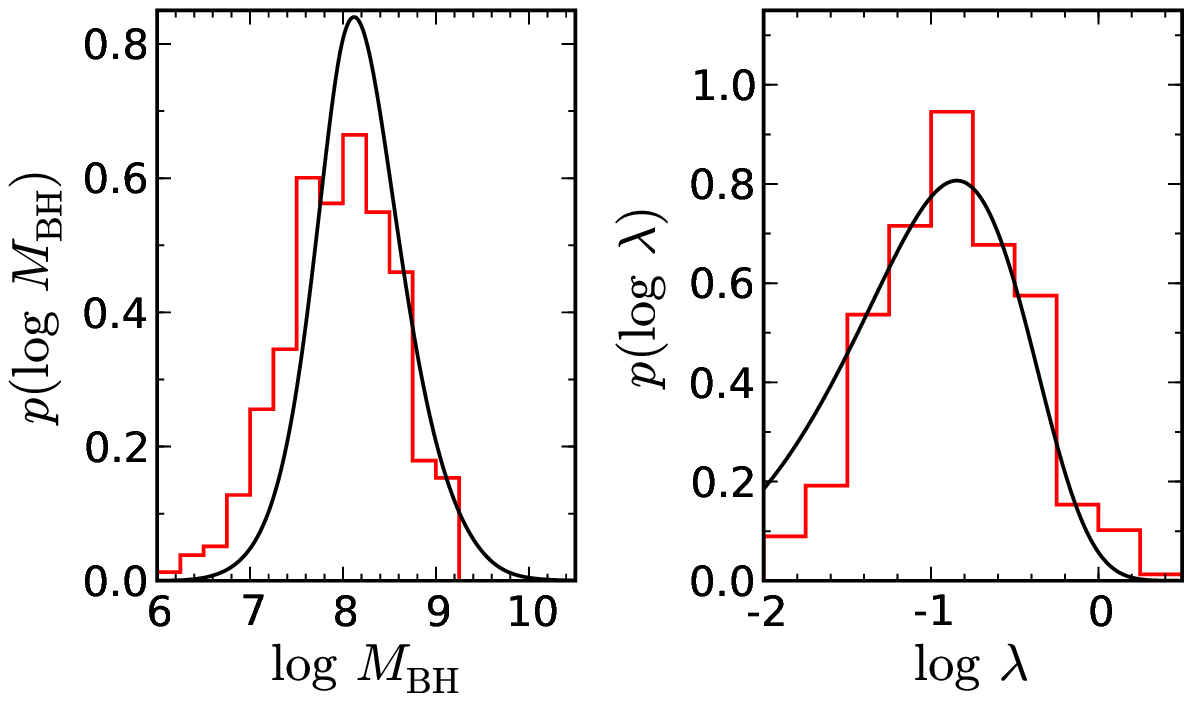}}

\put(5,0){\includegraphics[width=\wi]{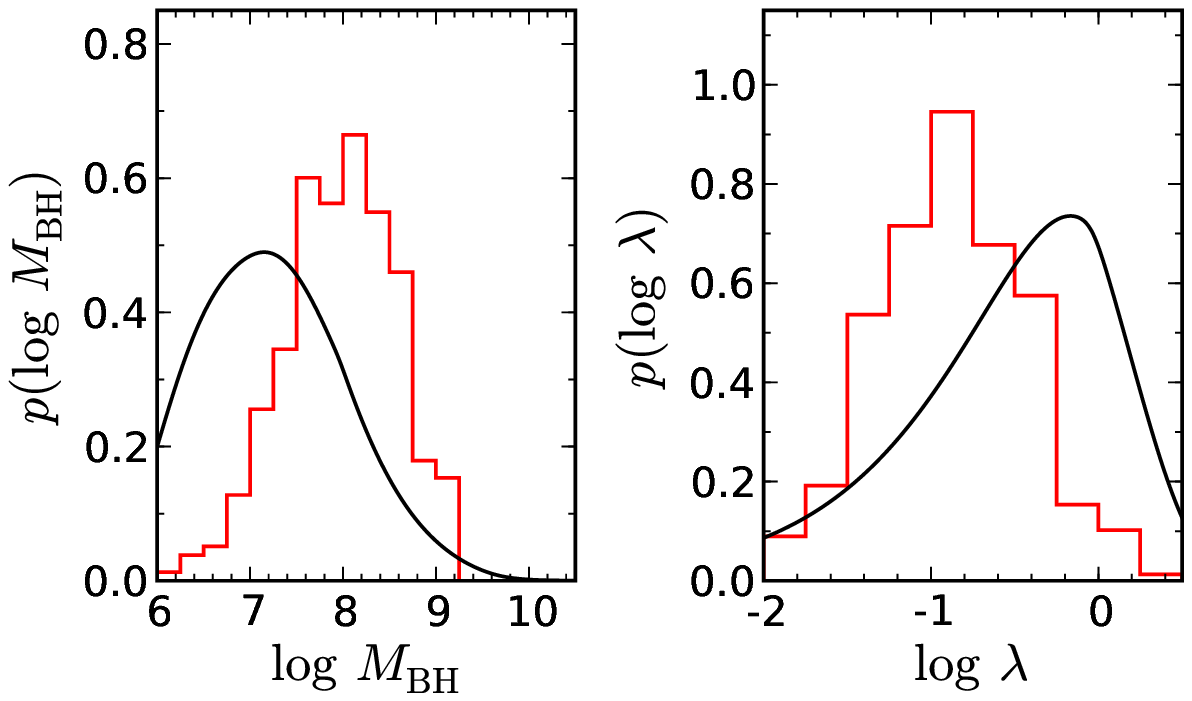}}
\put(63,0){\includegraphics[width=\wi]{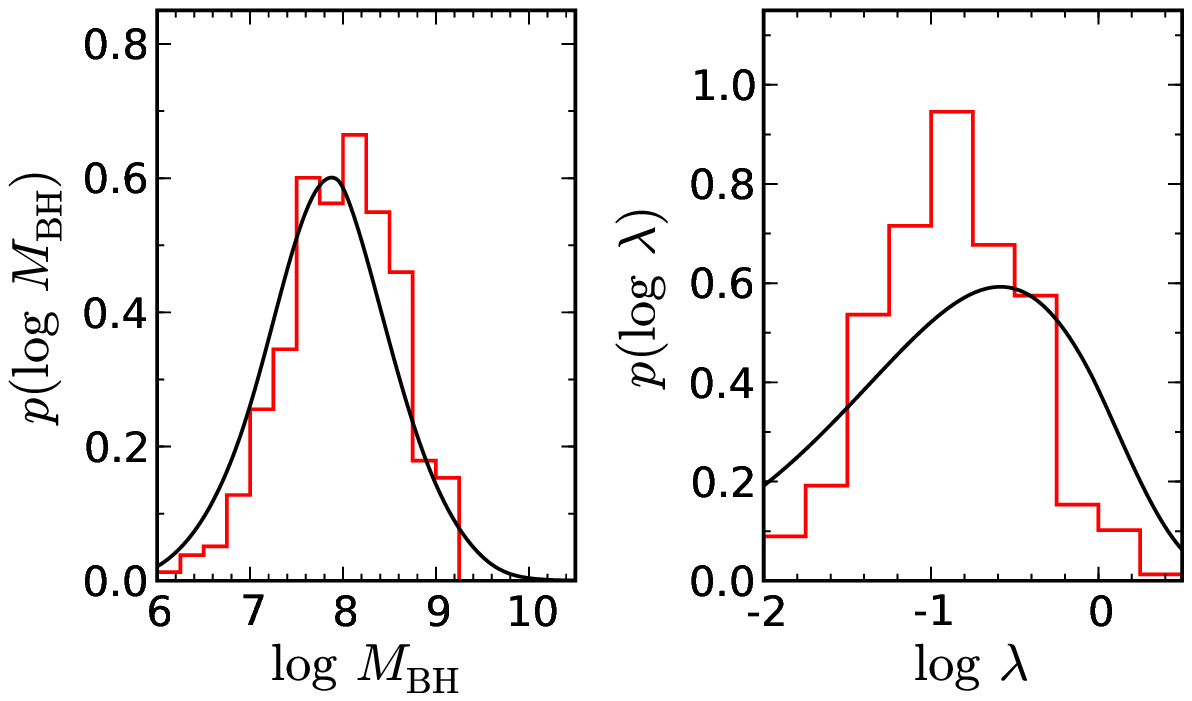}}
\put(121,0){\includegraphics[width=\wi]{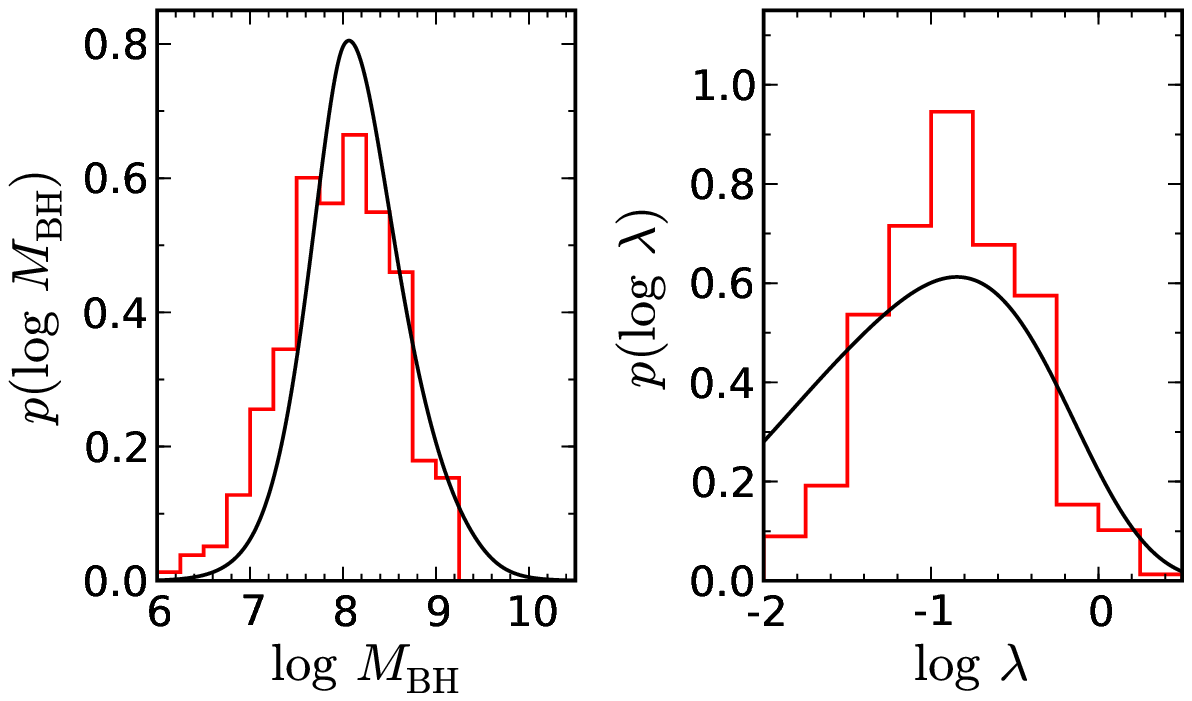}}
\end{picture}
\caption{Comparison of the expected distribution of \mbh and $\er$ (solid lines) with their distribution in the HES sample (red histogram) for 9 different combinations of the two free parameter $\alpha_\mathrm{BH}$ and $\alpha_\er$. The central panel is close to our best fit solution.}
\label{fig:distr_grid}
\end{figure*}

For illustration, in Fig.~\ref{fig:distr_grid} we compare these expected distributions with the observed ones within a grid of free parameter $\alpha_\mathrm{BH}$ and $\alpha_\er$. For a too steep BHMF the number of low mass objects is larger than observed, while for a too flat BHMF the number of low mass objects is lower than observed. A steep ERDF corresponds to a break of the ERDF close to the Eddington limit, thus more objects above the Eddington limit and less at low $\er$ are predicted, compared to the observations. For a too flat ERDF, the break needs to be at a low value of $\er$ and thus not enough objects close to the Eddington limit are predicted.

\begin{figure*}
\centering
	\includegraphics[width=15.3cm,angle=0]{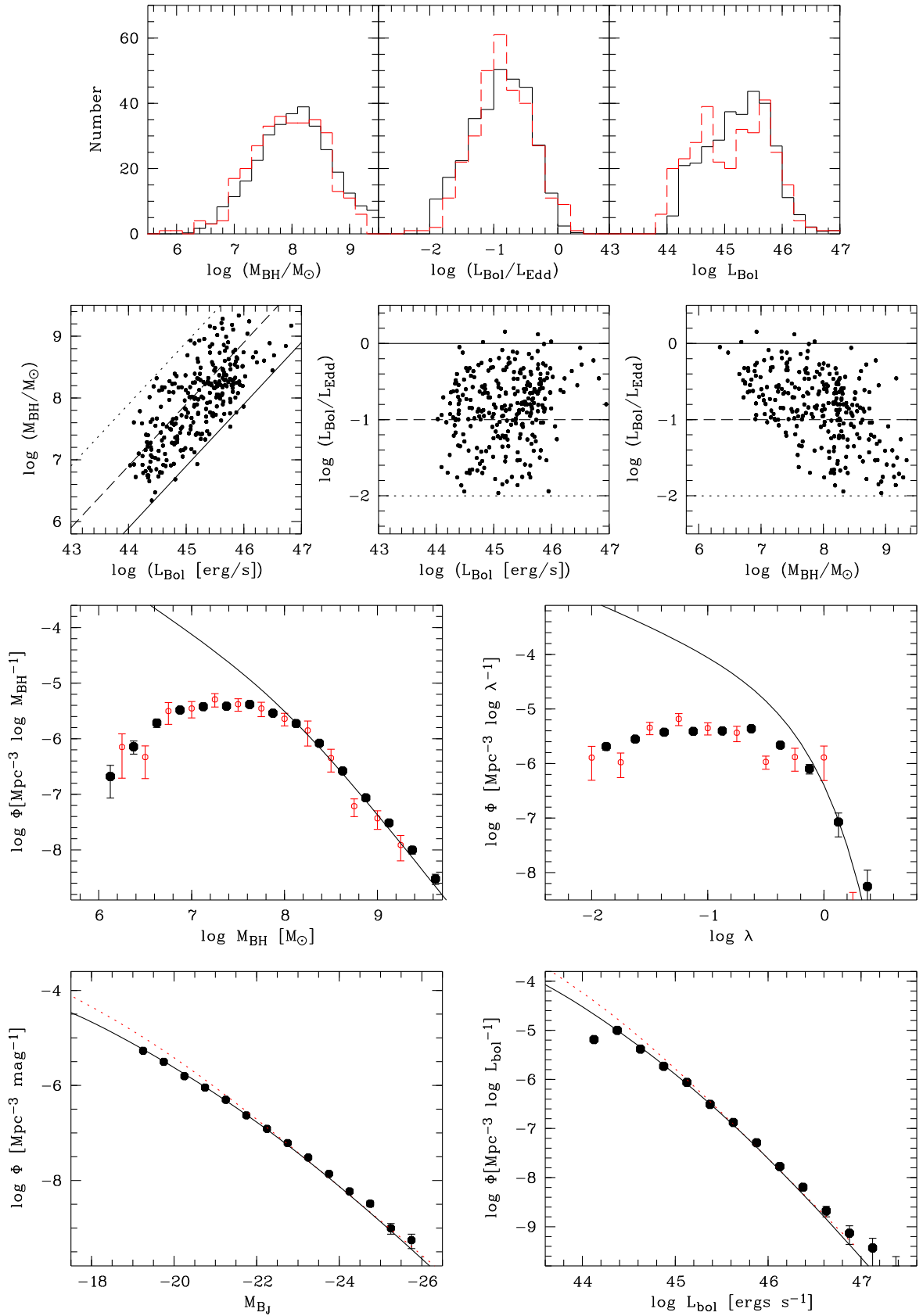}
\caption{Results of 10 Monte Carlo realizations for the best fit model with an assumed double power law with fixed high mass slope for the BHMF and a Schechter function parameterisation of the ERDF. Upper panels: Comparison of the distributions of \mbh, $\er$ and $L_\mathrm{bol}$ between simulated sample (black, solid histogram) and observed sample (red, dashed histogram). Middle panels: Same as Fig.~\ref{fig:Lbol}, but for one simulated sample. Lower panels: uncorrected BHMF and ERDF, $M_{B_J}$ luminosity function and bolometric luminosity function. The results for the simulated sample are shown as filled black points. The solid black line shows the true input function and the dotted lines show the best fit to the uncorrected distribution functions of the observed sample. The open red circles in the BHMF and ERDF plot indicate the individual bins for the observed uncorrected BHMF and ERDF with a restriction of $M_{B_J}<-19$ applied. }
\label{fig:bestmodel}
\end{figure*}

We also carried out Monte Carlo simulations for a grid of free parameters $\alpha_\mathrm{BH}$ and $\alpha_\er$, using the same assumptions as above, as well as for the best fit model of the maximum likelihood estimation. Here we proceeded as follows: First each AGN gets assigned a redshift, then its black hole mass is drawn from the assumed BHMF, and finally an Eddington ratio is drawn from the ERDF. From these values absolute and apparent $B_J$ magnitudes are computed, applying a bolometric correction. By means of the apparent magnitude $B_J$ it is decided if the object is selected by the survey or not, taking into account the flux-limit. 

We ran Monte Carlo simulations for a wide range of $\alpha_\mathrm{BH}$ and $\alpha_\er$ and found results consistent with what we discussed above and what is shown in Fig.~\ref{fig:distr_grid}. The Monte Carlo simulations are clearly able to discriminate between models that are consistent with the data and those that are not. The best matching solutions of the Monte Carlo simulations are consistent with the best fit from the maximum likelihood method, although 'best matching' is not as well defined in this case.

In Fig.~\ref{fig:bestmodel} we show the mean of 10 Monte Carlo realizations of this best fit model. We show the observed distributions for the sample for this model as well as the uncorrected BHMF and ERDF, as well as the $M_{B_J}$-LF and bolometric LF that would be determined from an 'observed' sample. To construct such an 'observed' sample we again limited the simulated sample to $M_{B_J}<-19$. In the middle panels of Fig.~\ref{fig:bestmodel}, we then compare these expected distribution functions with the uncorrected BHMF and ERDF determined with the same restriction applied (shown as open red symbols). The distributions as well as the constructed distribution functions are consistent with the observed distributions and distribution functions.
For models that are found to be not consistent with the observations based on the maximum likelihood approach, the Monte Carlo samples also provide a poor match to the observed distributions and distribution functions, and thus can also be rejected based on the Monte Carlo simulations.

These Monte Carlo simulations show that the observed distribution of objects between $L_\mathrm{Bol}$, \mbh and $\er$, as shown in the plots of Fig.~\ref{fig:Lbol}, are well understood by the underlying BHMF and ERDF and the selection function of the HES. These results do not qualitatively change using a different functional form for the BHMF or ERDF.

As a second model we again used a double power law, but included the high mass slope as an additional free parameter to be determined in the maximum likelihood fit. The result is shown as blue dashed dotted lines in Fig.~\ref{fig:trueBhmf}. The BHMF is highly consistent with the previous result, with a mild steepening of the high mass slope when this parameter is allowed to change in the fit.

Third, we also used the function given by Equation~\ref{eq:modschecht}, thus a modified Schechter function. 
The best fit result is consistent with the double power law fit over most of the mass range and only decreases stronger at the high mass end.
All three models are good representations of the observed data and therefore span the range of acceptable distribution functions. Formally, the modified Schechter function has the lowest value of $S$ and the highest probability both in the KS-test as well as in the $\chi^2$-test and we will use it in the following as our reference model.

Apart from the Schechter function for the ERDF, we additionally tested a log-normal distribution. This distribution function also provides a good representation of the data. In Table~\ref{tab:res} and Fig.~\ref{fig:trueBhmf} we give a model with a log-normal distribution for the ERDF and a modified Schechter function for the BHMF. While the BHMF is nearly unchanged, the ERDF deviates from the Schechter ERDF at the highest and lowest values, while being consistent over a wide range in between. When enforcing a turnover in the ERDF, using a log-normal distribution, the data are consistent with such a turnover at low $\er$ ($\log \er \approx -1.8$). However, there is no evidence for a turnover at higher $\er$, where the maximum in the \textit{observed} Eddington ratio distribution is present ($\log \er \approx -1$). 

The log-normal fit indicates rather a flattening of the ERDF at the low-$\er$ end then a real turnover, because it is cut off before the turnover, enforced by a log-normal fit, becomes evident. However, the low-$\er$ regime is dominated by high mass black holes. If there is a mass dependence in the ERDF and the ERDF flattens towards high \mbh, this would be most prominent at low $\er$. Such a flattening would also be consistent with \citet{Hopkins:2008}, who found evidence for a mass dependence in the ERDF of type~2 AGN, with a flatter low $\er$ slope at high \mbh.

We take into account the log-normal ERDF in the uncertainty range of the determination of the BHMF and ERDF. Formally it has a higher probability in the applied statistical tests than the Schechter function. However, as mentioned, the main deviation compared to the Schechter function is above the Eddington limit and close to the lower limit at $\er=0.01$. The number statistics in this regions are low and thus a clear discrimination between the two models is not possible. Thus, the Schechter function and log-normal distributions indicate the range of acceptable ERDFs.

\begin{figure*}
\centering
        \includegraphics[width=9cm]{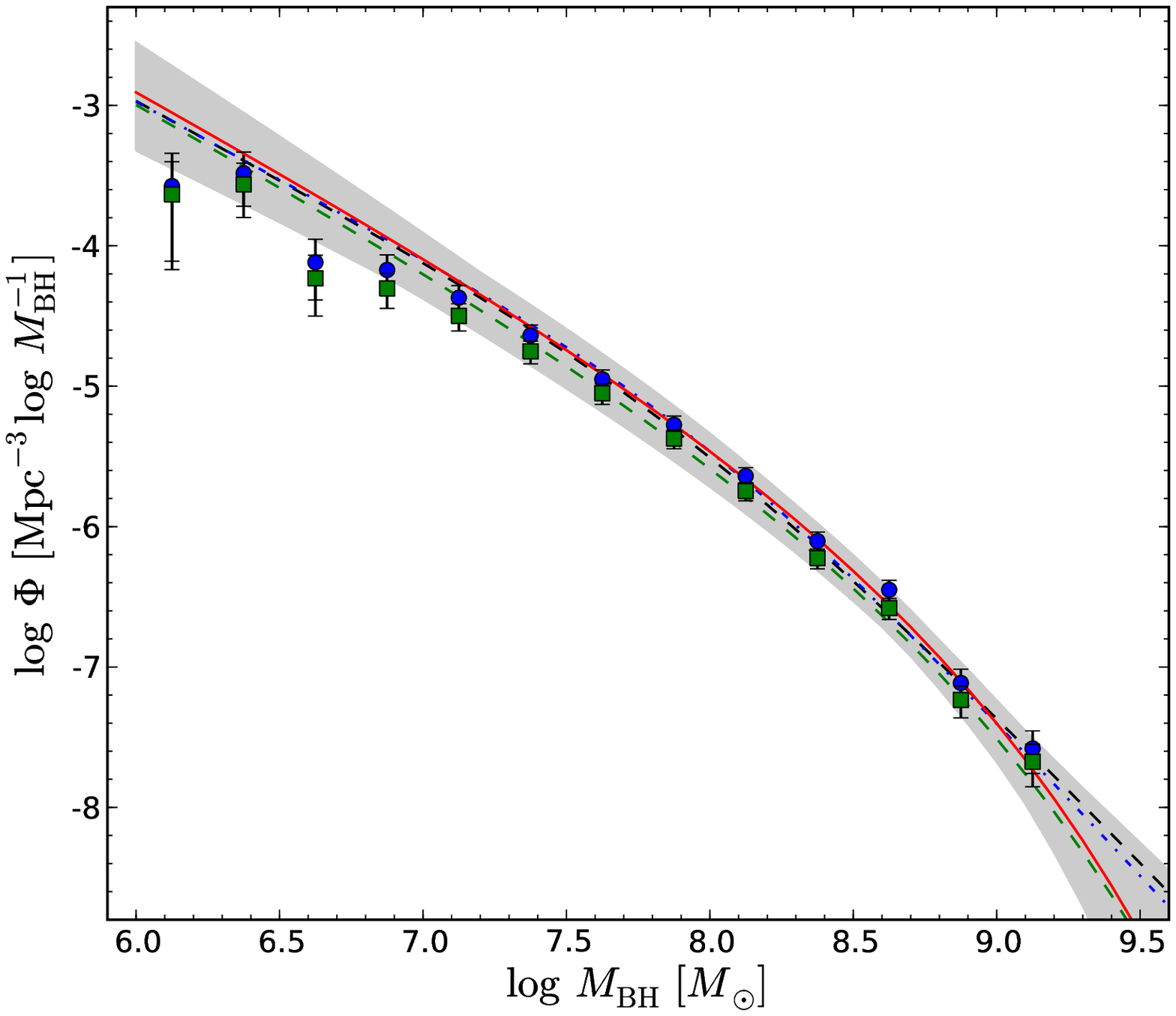}
        \includegraphics[width=9cm]{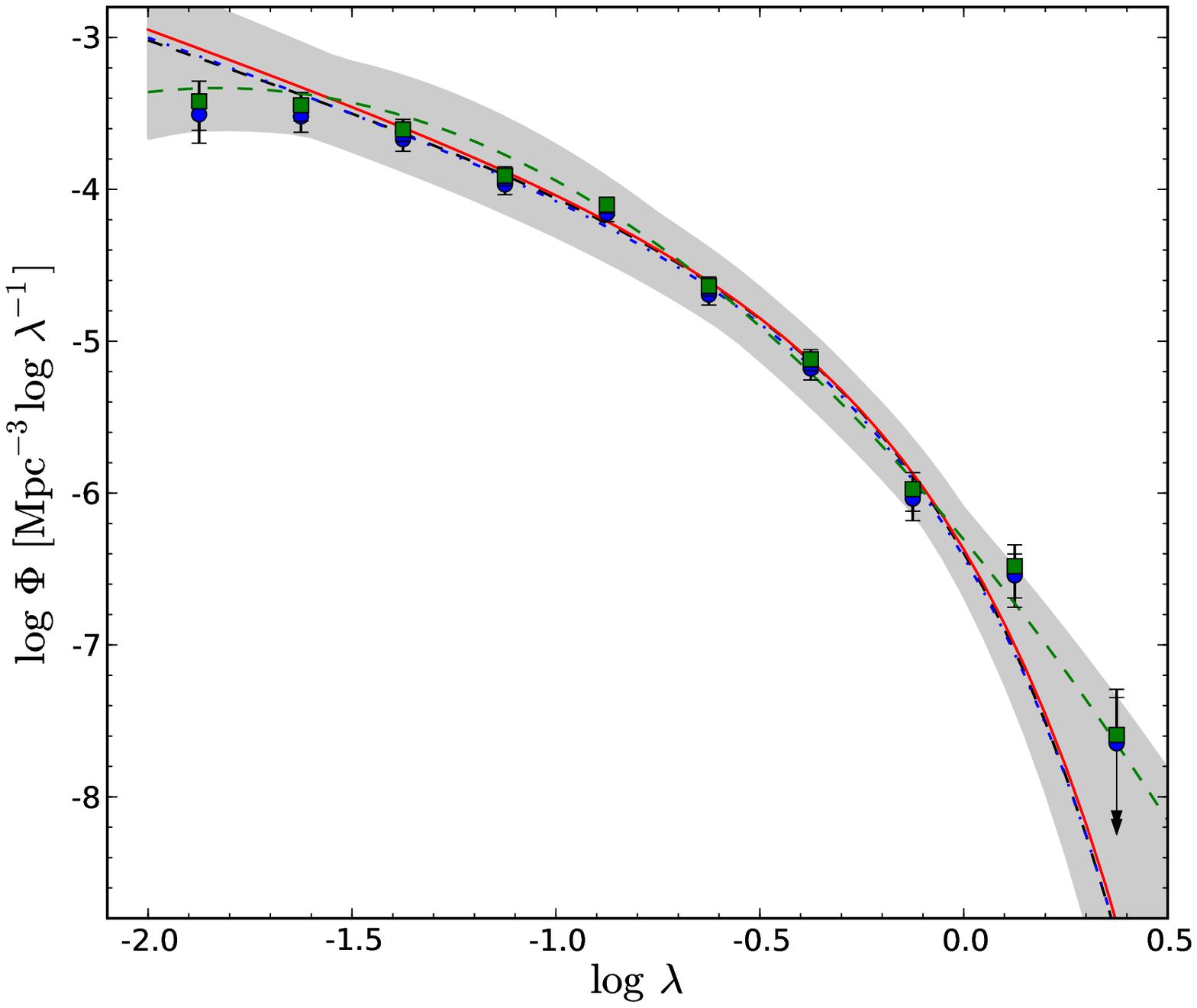}
\caption{Same as Fig.~\ref{fig:trueBhmf} with the constraints from the $1/V_\mathrm{max}$ method added (see Section~\ref{sec:intrinsic_va}). The binned results for the BHMF (left panel) were derived assuming the ERDF of the best fit solutions for the 4 models given in Table~\ref{tab:res}. Shown are only the two distinct models, the Schechter function (blue circles) and the log-normal distribution (green squares). Analogously, the binned results for the ERDF were derived assuming the BHMF of these 4 models. Shown are again only the two distinct models, the double power law (blue circles) and the modified Schechter function (green squares)}. The binned results for the different models are consistent with each other, as well as with the results of Section~\ref{sec:reconst}, shown as lines and by the shaded area. 
\label{fig:corBhmf}
\end{figure*}

We derived uncertainties in the BHMF and ERDF by randomly modifying the best fit parameters for each model and computing the likelihood function $S$. Using $\Delta S = S-S_\mathrm{min}$ for each random realization, we converted $\Delta S$ into confidence values assuming a $\chi^2$ distribution \citep{Lampton:1976, Press:1992}. For all models within a certain confidence interval the BHMF and ERDF is computed and these functions then span the confidence range of the two distribution functions. The total uncertainty of the BHMF or ERDF is then the sum of the confidence ranges of the individual models. In Fig.~\ref{fig:trueBhmf}, we show this sum of the $1\sigma$ confidence values for the two distribution functions as the gray shaded areas.

So far, we assumed the estimated black hole mass to be equal to the true black hole mass. However, this is probably an oversimplification. It is known that there is a considerable uncertainty in \mbh estimates using the virial method, probably of order $0.4$~dex \citep{Vestergaard:2006}. Accounting for this uncertainty might change the reconstructed BHMF and ERDF in shape as well as in normalisation. We will investigate this important point in detail in future work.

\subsection{BHMF and ERDF from the $1/V_{\mathrm{max}}$ method} \label{sec:intrinsic_va}
As mentioned in Section~\ref{sec:bhmf}, there is also a different approach to determine the intrinsic BHMF and ERDF, namely using the $1/V_{\mathrm{max}}$ method, but directly accounting for the selection effects in terms of black hole mass or Eddington ratio completeness imposed on the sample by the AGN luminosity selection. In this case, the BHMF and ERDF cannot be determined jointly. When using the $1/V_{\mathrm{max}}$ method the selection effects need to be accounted for in the determination of the accessible volume of the individual AGN, given by:
 \begin{equation} 
	V_\mathrm{max} 
	= \int_{z_\mathrm{min}}^{z_\mathrm{max}} \Omega_{\mathrm{eff}} \frac{\dd V}{\dd z} \dd z  	\label{eq:vmax} \ ,
\end{equation}
where $\Omega_{\mathrm{eff}}$ is the effective survey area as a function of apparent magnitude, thus the selection function for our flux-limited survey, depending on $z$, \mbh and $\er$, is $\Omega_{\mathrm{eff}}(m)=\Omega_{\mathrm{eff}}(L,z)=\Omega_{\mathrm{eff}}(M_\bullet,\er,z)$. While for the determination of the luminosity function the proper selection function to compute $V_\mathrm{max}$ is given by $\Omega_{\mathrm{eff}}(L,z)$, using it for the determination of the BHMF (as we did in Section~\ref{sec:bhmf}) will lead to the presence of sample selection effects, and thus to the observed underestimation of the space density at low mass.

This bias on the determined BHMF can be avoided by using a black hole mass selection function, given by:
\begin{equation}
\Omega_{\mathrm{eff}}(M_\bullet,z)= \int^\infty_{\er_\mathrm{min}} P_\er(\er) \Omega_{\mathrm{eff}}(M_\bullet,\er,z) \dd \log \er \ , \label{eq:effarea_bh}
\end{equation} 
where $P_\er(\er)$ is the normalised ERDF, given by Equation~\ref{eq:P_er}. However, this approach requires knowledge of the ERDF as prior information, which is not present a priori.

Likewise, the ERDF can be derived in an unbiased way by using the Eddington ratio selection function for the survey, given by:
\begin{equation}
\Omega_{\mathrm{eff}}(\er,z)= \int^\infty_{M_\mathrm{min}} P_\mathrm{BH}(M_\bullet) \Omega_{\mathrm{eff}}(M_\bullet,\er,z) \dd \log \mbh \ , \label{eq:effarea_er}
\end{equation} 
where $P_\mathrm{BH}(\mbh)$ is the normalised BHMF, similar to $P_\er(\er)$, defined by:
\begin{equation}
P_\mathrm{BH}(M_\bullet)=\frac{\Phi_\bullet(M_\bullet)}{  \int \Phi_\bullet(M_\bullet)\,\dd\log M_\bullet }\ . \label{eq:P_bh}
\end{equation} 
This reqires knowledge of the BHMF, which is also unknown beforehand. Thus this approach is usually not feasible for the determination of the intrinsic BHMF and ERDF directly from the data.

However, this approach has the advantage that no prior assumptions on the shape on the ERDF are required for their determination, once we fixed the assumed BHMF. The same is equally true for the determination of the BHMF. The only necessary information beforehand is on the shape of the ERDF. The problem is that one distribution function needs to be known to determine the other one.

Nevertheless, first we can use it as a consistency test, constructing the BHMF from the constraints on the ERDF from Section~\ref{sec:reconst} and vice versa. The resulting binned BHMFs and ERDFs using the 4 best fit models are shown as filled symbols in Fig.~\ref{fig:corBhmf} together with the best fit solutions to the active BHMF and the ERDF, as determined in Section~\ref{sec:reconst}. These binned BHMFs as well as the binned ERDFs for all 4 models are fully consistent with our previous constraints and also consistent with each other.

On the other hand, this approach is useful to verify the assumptions on the shape of the distribution functions used in Section~\ref{sec:reconst}. This is especially worthwhile for the ERDF, because the shape of the BHMF is relatively well determined at the high mass end, with the main uncertainty in the low mass slope, while the shape of the ERDF is poorly determined. Therefore, we assumed the double power law with fixed high mass slope parameterisation of the BHMF. As shown above, the shape of the binned ERDF is consistent for all four assumed BHMFs, based on the 4 best fit models. Thus it is justified to use one of these for the investigation of the ERDF shape.

We again fix the break of the double power law and thus the only free parameter left is the low mass slope $\alpha_\mathrm{BH}$. We determined the Eddington ratio selection function, using Equation~\ref{eq:effarea_er} for a variety of values for $\alpha_\mathrm{BH}$, covering the whole range of acceptable values. We use $\alpha_\mathrm{BH}=-0.7$ as lower limit, taken from the uncorrected BHMF, and  $\alpha_\mathrm{BH}=-3.2$ as upper limit, corresponding to a single power law BHMF. The fitting results on these ERDFs with a Schechter function are given in Table~\ref{tab:cor}. While the normalisation of these ERDFs changes significantly for different assumed values of $\alpha_\mathrm{BH}$, the shape is not strongly affected and is consistent with our previous constraints thoughout the whole range. In particular, the ERDF is well described by a Schechter function. While there is an indication for a flattening at the low $\er$ end, no indication for a real turnoff of the ERDF is present, as also shown in the right panel of Fig.~\ref{fig:corBhmf}. A log-normal distribution is also appropriate, but needs to be cut off close to the maximum of the distribution. Thus it does not indicate a turnover, but only a flattening of the ERDF.
We again want to emphasise that no prior assumptions on the ERDF are used here, we just modified the selection function using an assumed BHMF over a wide range of possible parameters. This strongly confirms our previous results for the shape of the ERDF, in that it shows that a Schechter function provides a good representation of the data.

\begin{table}
\label{tab:cor}
\caption{Fitting results for the ERDF, determined using an appropriate Eddington ratio selection function, assuming different values for the low mass slope $\alpha_\mathrm{BH}$ of the BHMF.}
\centering
\begin{tabular}{rrrr} \hline\hline \noalign{\smallskip}
$\alpha_\mathrm{BH}$ & $\alpha_\er$ & $\log \er_\ast$ & $\chi^2$/d.o.f.\\ 
\hline \noalign{\smallskip}
$-$0.7 & $-$1.51 & $-$0.63 & 1.81 \\ \noalign{\smallskip}
$-$1.2 & $-$1.50 & $-$0.66 & 1.77 \\ \noalign{\smallskip}
$-$1.7 & $-$1.51 & $-$0.69 & 1.73 \\ \noalign{\smallskip}
$-$2.0 & $-$1.55 & $-$0.70 & 1.73 \\ \noalign{\smallskip}
$-$2.2 & $-$1.59 & $-$0.71 & 1.75 \\ \noalign{\smallskip}
$-$2.2 & $-$1.59 & $-$0.71 & 1.75 \\ \noalign{\smallskip}
$-$2.7 & $-$1.79 & $-$0.73 & 1.92 \\ \noalign{\smallskip}
$-$3.0 & $-$1.94 & $-$0.72 & 2.02 \\ \noalign{\smallskip}
$-$3.2 & $-$2.04 & $-$0.71 & 2.02 \\ \noalign{\smallskip}
\hline
\end{tabular} 
\end{table}

\begin{figure*}
\centering
\includegraphics[height=17cm,angle=-90]{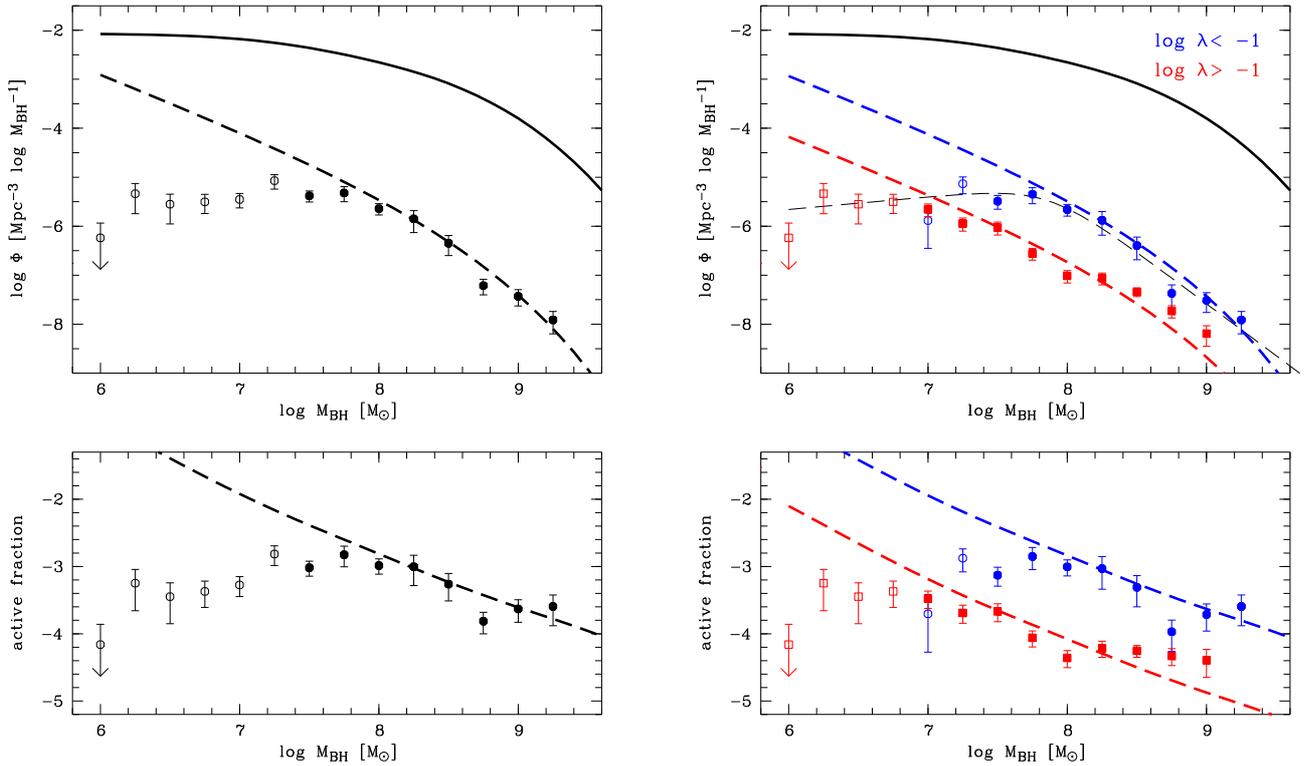}
\caption{Comparison of our active black hole mass function with the inactive  BHMF of \citet{Marconi:2004} (solid line in upper panels). The circles show the uncorrected binned data, where open symbols correspond to bins that suffer from selection effects.
In the lower panels the active fraction of black holes in the local universe is shown in logarithmic units. In the left panel the active BHMF and active fraction for the whole sample are shown. In the right panel the active BHMF and the corresponding active fraction are shown for two Eddington ratio bins (above and below $\log \er = -1$). The BHMF and active fraction for the best fit model of the intrinsic BHMF and ERDF are shown as dashed lines, in the left panel for the whole sample and in the right panel for the two $\er$ bin subsamples. There is  a decrease in the active fraction with increasing black hole mass, in agreement with the cosmic downsizing picture. This decrease is already visible in the high~$\er$ subsample (red squares).}
\label{fig:acf}
\end{figure*}

\section{Discussion} \label{sec:discus}

\subsection{Active fraction of local black holes} \label{sec:af}

For a census of active black holes, the derived mass function of active black holes should be compared to the local mass function of quiescent black holes. Because the number of dynamically measured black hole masses is still very low and the sample is inhomogeneous, the quiescent black hole mass function has to rely on the known \mbh - bulge property relations, thus converting galaxy luminosity or velocity functions into a black hole mass function. This approach has been used by several authors to derive a local BHMF \citep[e.g.][]{Salucci:1999,Yu:2002,Shankar:2004,Marconi:2004}. However, there is still some uncertainty in the estimation \citep{Tundo:2007}. We compare our active BHMF to the BHMF presented by \citet{Marconi:2004}, shown as the solid line in Fig.~\ref{fig:acf}. Our best fit model of the reconstructed active BHMF, derived above, is indicated as dashed line in Fig.~\ref{fig:acf}.

At this point we need to recall that our operational definition of 'active' black holes only includes type~1 AGN.
We are not able to distinguish between a true quiescent black hole and an AGN not selected due to obscuration. By dividing our active BHMF by the quiescent BHMF we thus get the fraction of black holes in an active stage, not hidden to our survey by obscuration, and thus a lower limit to the true active fraction.

The lower panels in Fig.~\ref{fig:acf} show the fraction of local black holes in an active stage as a function of the black hole mass, thus the black hole duty cycle.

As circles we give the active fraction, or duty cycle, derived from the binned uncorrected BHMF, presented in Section~\ref{sec:bhmf}, where open symbols indicate bins that are affected by sample censorship. The estimate of the active fraction for the intrinsic BHMF is shown as dashed line, thus showing the intrinsic underlying black hole duty cycle. A clear decrease of the active fraction with increasing \mbh is visible, being close to a power law with slope $\alpha_\mathrm{AF} \approx -0.86$ over the whole covered mass range.

Using a very different approach, \citet{Shankar:2007} predicted the black hole duty cycle. They used
simple black hole growth models, based on the local quiescent BHMF and the bolometric AGN luminosity function. They made the simplified assumption of a single constant accretion rate, in contrast to the wide accretion rate distribution we assumed. Their active fraction also refers to the whole AGN population, while we are restricted to type~1 AGN. Nevertheless, when comparing their low $z$ duty cycle with our results, we find an excellent agreement between both. However, taking into account the large differences between the simple model of \citet{Shankar:2007} and our empirical determination, this agreement might even be a coincidence.

In the right panels of Fig.~\ref{fig:acf} we split our sample into two subsamples, based on the Eddington ratio $\er$, at $\log \er=-1$. For both subsamples we computed the uncorrected BHMF and the active fraction. 
The uncorrected BHMF and the uncorrected active fraction for the low $\er$ subsample are shown as blue circles, while the blue dashed line shows the active fraction derived from the reconstructed BHMF (best fit modified Schechter function). Incompleteness sets in around $10^8\,M_\odot$ and is dominant below  $10^7\,M_\odot$, therefore no information on the behaviour of the active fraction can be gained from these low $\er$ black holes.  

The high $\er$ subsample is shown as red squares, while the red dashed line shows the active fraction derived from the reconstructed BHMF, with the normalisation derived from the fraction of objects above $\log \er=-1$. The subsample is almost complete up to $\sim10^7\,M_\odot$, where the low $\er$ subsample is already heavily incomplete. Above $10^7\,M_\odot$ the binned active fraction is in good agreement with the reconstructed intrinsic active fraction. This provides a consistency test for the reconstructed BHMF and ERDF estimate.
But even without this comparison there is a clear trend present for the high $\er$ subsample with a decrease of the active fraction with increasing black hole mass, directly verifying our previous result from the uncorrected binned data. Thus, far more low mass black holes in the local universe are in an active state than high mass black holes.

This result is in general agreement with the picture of anti-hierarchical growth of black holes \citep[e.g.][]{Merloni:2004, Merloni:2008}, where the most massive black holes grew at early cosmic times and are preferentially in a less active stage in the present universe, and at present mainly smaller mass black holes grow at a significant rate, also known as cosmic downsizing. Our results strongly support this anti-hierarchical black hole growth scenario.
This is in general agreement with previous findings on low redshift AGN \citep{Heckman:2004,Greene:2007a,Goulding:2010} that also report a decrease of the active fraction for the most massive black holes, as well as with results at higher redshifts \citep{Vestergaard:2009}.

\subsection{The active black hole mass density} \label{sec:bhmassdens}
We now want to estimate the black hole mass density of active type~1 AGN in the local universe
\begin{equation}
\rho_\mathrm{act} =\int_{M_\mathrm{min}}^\infty M_\bullet \phi(M_\bullet) \mathrm{d}M_\bullet \ ,
\end{equation}
with $M_\mathrm{min}=10^6 M_\odot$, using our results for the active BHMF.
A lower limit on the local mass density of active black holes is given by the BHMF without a correction for sample censorship. We derived a lower limit of $\rho_\mathrm{act}=277\,M_\odot\, \mathrm{Mpc}^{-3}$. Using our reconstructed BHMF, the local mass density of active black holes with $\log \er > -2$ is then $\rho_\mathrm{act}\approx 1700 \, M_\odot\, \mathrm{Mpc}^{-3}$, a factor of $6$ higher then derived from the uncorrected active BHMF. The results for the individual models are given in Table~\ref{tab:res}. 

The observational estimate of the integrated mass density of the total black hole population in the local universe is $\rho_\mathrm{tot}=(3.2-5.4)\times 10^{5} \, M_\odot\, \mathrm{Mpc}^{-3}$ \citep{Shankar:2007,Graham:2007,Yu:2008}. Using a value of $4.6\times 10^{5} \, M_\odot\, \mathrm{Mpc}^{-3}$, as presented by \citet{Marconi:2004}, results in a fraction of $\sim 4\times 10^{-3}$ of the black hole mass that is currently actively accreting at a rate larger that 1\% of the Eddington limit ($\sim 6\times 10^{-4}$ for the uncorrected BHMF).

\subsection{Comparison with other surveys}
\citet{Greene:2007a} presented a determination of the active black hole mass function for $z<0.352$, using the SDSS DR4 main galaxy sample as well as the QSO sample. They constructed their sample based on spectroscopic confirmation of broad H$\alpha$ lines, ending up with 8728 objects. For these they computed black hole masses from the H$\alpha$ FWHM and line luminosity. 

\begin{figure}
\centering
\resizebox{\hsize}{!}{\includegraphics[angle=-90,clip]{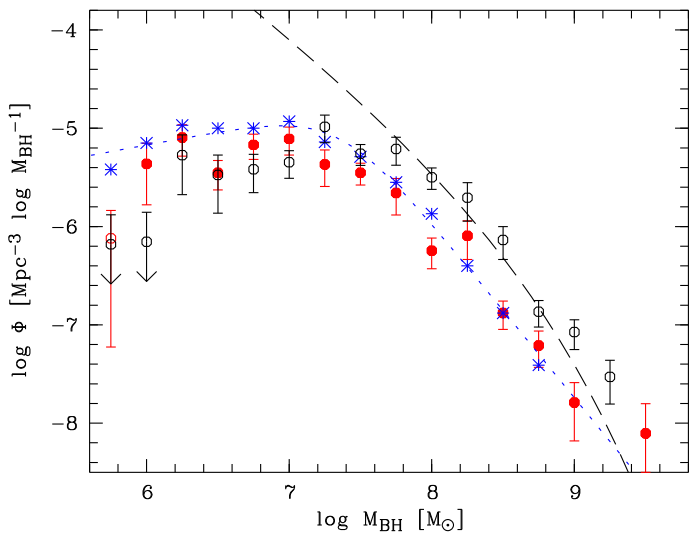}}
\caption{Comparison of the local BHMF of the HES with the BHMF presented in \citet{Greene:2009}. The blue asterisks and the blue dotted line show the BHMF from \citet{Greene:2007a} (corrected in \citet{Greene:2009}). The open, black circles show our BHMF, not corrected for evolution and sample censorship, while the filled, red circles show our BHMF, with the black hole mass estimated as in \citet{Greene:2007a}. The dashed line indicates our reconstructed BHMF for reference.}
\label{fig:bhmf_gh07}
\end{figure}

As already mentioned in Paper~I, an error has been discovered in the determination of the $V_\mathrm{max}$ values in the work of \citet{Greene:2007a} (J. Greene, private communication), resulting into an erroneous  luminosity function as well as BHMF. This error has recently been resolved \citep{Greene:2009}. Thus we caution not to use the original active BHMF from \citet{Greene:2007a}. In Fig.~\ref{fig:bhmf_gh07} the active BHMF by \citet{Greene:2009} from their SDSS sample is shown as blue asterisks. 

\citet{Greene:2007a} have not taken into account the selection effects caused by the use of the luminosity selection function and thus underestimate the number of active black holes at low masses, due to the discussed sample censorship. They also do not correct for evolution within their $z$ range. However, a direct comparison with the mass function from \citet{Greene:2009} can be made using our BHMF, without correction for evolution and sample censorship. 

For consistency, we also re-estimated the black hole masses of our sample, using the same formula as \citet{Greene:2007a}, using H$\alpha$ FWHM and H$\alpha$ line luminosity. For our sample, the black hole mass distribution is shifted by 0.54~dex towards lower mass in the mean. Compared to the FWHM based \mbh this shift is 0.42~dex, thus $\sim0.1$~dex can be attributed to the difference between the FWHM and \ldisp based \mbh. The main reason for the remaining difference originates from a different relation of H$\alpha$ luminosity to $L_{5100}$ found for our sample compared to the one given in \citet{Greene:2005}. This difference leads to an offset of 0.31~dex. The remaining offset can be attributed to the different $R_\mathrm{BLR}-L$ scaling relation as well as to scatter in the relation between the FWHMs, as shown in Fig.~\ref{fig:reg}.

The resulting BHMF of the HES is shown as filled circles in Fig.~\ref{fig:bhmf_gh07}. Both BHMFs are fully consistent with each other, especially at the high mass end, where different survey selection effects are not important. At the low mass end the SDSS BHMF seems to exhibit similar survey selection effects as our HES sample, resulting in a consistent uncorrected BHMF, even at the biased low mass end. 

\begin{figure}
\centering
\resizebox{\hsize}{!}{\includegraphics[angle=-90,clip]{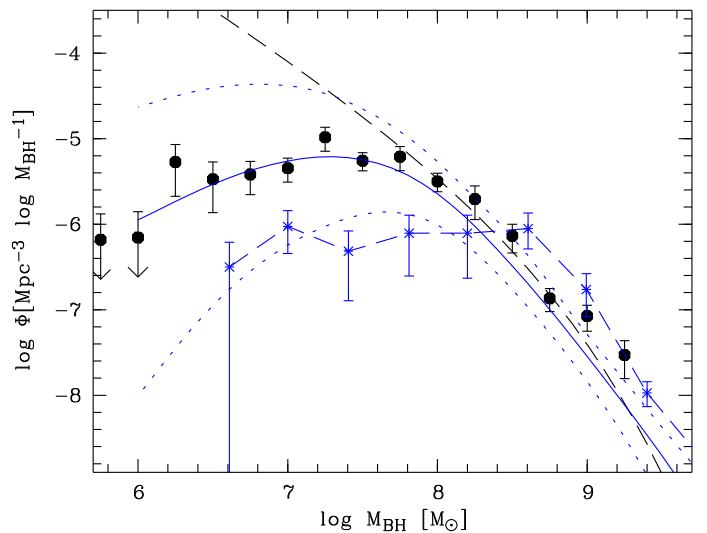}}
\caption{Comparison of our local active BHMF (filled circles for uncorrected and dashed line for intrinsic BHMF) with the BHMF of the BQS, as determined directly as binned estimate by \citet{Vestergaard:2009}(blue asterisks) and as determined by a Bayesian approach by \citet{Kelly:2008b}(blue solid line for median and dashed lines for $1\sigma$ uncertainty).}
\label{fig:bhmf_bqs}
\end{figure}

Recently, \citet{Vestergaard:2009} presented the binned local active BHMF of the Bright Quasar Survey \citep[BQS,][]{Schmidt:1983}, in the redshift interval $z=0 - 0.5$. In Fig.~\ref{fig:bhmf_bqs} we compare their derived BHMF with our binned BHMF, not corrected for evolution and sample censorship (filled black circles), and our reconstructed intrinsic BHMF (dashed black line). We also show the local BHMF of the BQS as blue solid line, but determined using a Bayesian approach \citep{Kelly:2008b}.

The most direct comparison between the BQS and the HES is with the binned estimates. At the high mass end, both binned estimates are in reasonable agreement. However, the BQS does not cover exactly the same redshift range as our HES sample. This might also cause some difference between both BHMFs, due to evolution of the BHMF, which has the largest effect at the high mass end. Because the BQS is not as deep as the HES, incompleteness sets in at higher \mbh in the binned BHMF. Also it is known that the BQS suffers from both incompleteness \citep{Goldschmidt:1992,Koehler:1997} as well as overcompleteness \citep{Wisotzki:2000c}. Thus, the HES is superior to the BQS for a determination of the local active BHMF.

Recently, \citet{Kelly:2008b} presented a determination of the active BHMF from the BQS using a Bayesian method, taking also into account scatter in \mbh and accounting for black holes below the flux limit of the survey. Their approach aims at correcting their BHMF for sample selection effects caused by the flux-limit, as we did in Section~\ref{sec:intrinsic}. However, they modeled the BHMF with a combination of Gaussian functions and also enforced a log normal distribution for the ERDF, while we mainly used a Schechter function description without a specific maximum and with a high fraction of objects at low $\er$.
In Fig.~\ref{fig:bhmf_bqs} we compare their posterior median BHMF (blue solid line) with our intrinsic BHMF (black dashed line). While both mass functions are consistent at the high mass end, there is a clear disagreement at the low mass end. Their BHMF is rather consistent with our uncorrected BHMF. We speculate that the reason for this disagreement might lie in the different assumptions on the shape of the BHMF and ERDF. This emphasises the importance of the assumed ERDF for the determination of the underlying BHMF. An important constraint on the ERDF is provided by the condition to recover the observed luminosity function as a convolution of BHMF and ERDF, as we have ensured by construction.

So far, little observational results exist on the distribution function of Eddington ratios from AGN surveys. \citet{Yu:2005} used a sample of type~2 AGN from the SDSS \citep{Kauffmann:2003,Heckman:2004} to determine the ERDF. Their results have recently been compiled by \citet{Hopkins:2008}. They also fitted the ERDF by a Schechter function and found an average slope of $\alpha_\er \approx -1.6$. Our constraints on the local type~1 ERDF presented here are consistent with this average slope of the ERDF of type~2 AGN. This might indicate a similar accretion behaviour of type~1 and type~2 AGN, as expected from the standard unification model \citep[e.g.][]{Antonucci:1993}.

\begin{figure}
\centering
\resizebox{\hsize}{!}{\includegraphics[angle=-90,clip]{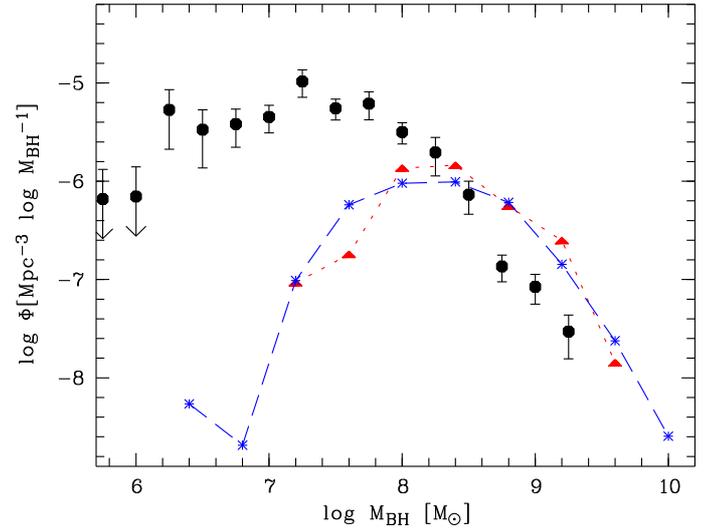}}
\caption{Comparison of our uncorrected $z\leq0.3$ BHMF (filled circles) with the active BHMF within $0.3\leq z\leq0.7$ from the SDSS \citep{Vestergaard:2008}, shown as the blue dashed line and asterisks, and the active BHMF of the LBQS within $0.2\leq z\leq0.5$ \citep{Vestergaard:2009}, shown as the red dotted line and triangles.}
\label{fig:bhmf_v}
\end{figure} 
\subsection{Evolution of the active fraction} \label{sec:evoaf}
\citet{Vestergaard:2008} presented a determination of the BHMF in the redshift range $0.3\leq z\leq5$. They used a well-defined, homogeneous sample of 15180 quasars from the SDSS DR3, already used by \citet{Richards:2006a} for the construction of the luminosity function. They found a high-mass decline with constant slope $\beta \approx -3.3$ at all epochs. Our high-mass slope of $\beta=-3.1$ for $z\leq0.3$ (when not corrected for evolution) is consistent with their higher-$z$ result within the uncertainties.

We compare our $z\leq0.3$ BHMF, not corrected for evolution within the $z$-bin and for sample censorship, with the lowest redshift bin ($0.3\leq z\leq0.68$) BHMF of \citet{Vestergaard:2008}, shown in Fig.~\ref{fig:bhmf_v}. We also show the active BHMF of the Large Bright Quasar Survey \citep[LBQS; e.g. ][]{Hewett:2001} in the redshift bin $z=0.2 -0.5$ \citep{Vestergaard:2009} as triangles. Both SDSS and LBQS BHMFs are in general agreement, even though they do not cover exactly the same redshift range. The decline of the space density at the lowest \mbh in both BHMFs is mainly due to incompleteness in this mass range in the SDSS QSO sample as well as in the LBQS QSO sample. At the high mass end the BHMF shows a similar slope but a larger space density than our HES BHMF. This seems to indicate evolution of the BHMF between these redshift bins. Because the mass function of the total supermassive black hole population will only decrease at the high mass end toward higher $z$, we can use the local quiescent BHMF as an upper limit for the mass function at $0.3\leq z\leq0.68$. This then implies an increase of the active fraction at the high mass end towards higher redshift, exactly as would be expected in the cosmic downsizing scenario. 
Thus, the number of the most massive black holes being in an active stage in the present universe seems to be lower than at earlier cosmic epochs.

\section{Conclusions}
We have presented a study of the low-redshift active black hole population, residing in broad-line active galactic nuclei. We estimated black hole masses and Eddington ratios, and from it estimated the local active black hole mass function and the Eddington ratio distribution function. Our sample was drawn from the Hamburg/ESO Survey and contains 329 quasars and Seyfert~1 galaxies with $z<0.3$, selected from surveying almost 7000~deg$^2$ in the southern sky.

We estimated black hole masses from single-epoch spectra, measuring the line dispersion of the broad H$\beta$ line and the continuum luminosity at 5100\,\AA{}\, $L_{5100}$, using the common virial method. We took care to avoid contamination of the line measurement by neighbouring emission lines and roughly estimated the degree of host galaxy contribution to $L_{5100}$. This has been found to be negligible for the most luminous AGN and not dominant even at the low luminosity end of our sample. We applied a rough statistical correction to the continuum luminosities to take into account the host contribution.
The bolometric luminosity and thus the Eddington ratio $\er$, has been estimated from $L_{5100}$.

The observed black hole masses cover a range $10^6 - 2\cdot10^9 \,M_\odot$ and the Eddington ratio is roughly confined between $0.01 -1$. The observed distributions of these quantities are understood by the underlying distribution functions of black hole mass and Eddington ratio, in combination with the survey selection function, as we explicitly demonstrated by Monte Carlo simulations.

We made an attempt to determine these two distribution functions in an unbiased way. First of all, when we want to determine the \textit{active} BHMF, we have to make clear what we mean by an \textit{active} black hole, due to the wide distribution of accretion rates. We used a lower Eddington ratio cut of $\log \er = -2$, in agreement with the observed range of Eddington ratios. Using a different cut for $\er$ will preserve the shape of the BHMF, but change their normalisation, due to our assumption of an uncorrelated BHMF and ERDF. This is already shown in the left panel of Fig.\ref{fig:acf}. The normalisation and therefore the space density clearly depend of the chosen definition of an active black hole.

Next we have to be aware that our sample is selected on AGN luminosity, not on black hole mass.
Therefore, we have to make sure that we properly account for active black holes below the flux limit of the survey. We presented a method that determines the active BHMF as well as the ERDF at the same time, by a maximum likelihood fit. Here, the bivariate probability distribution of black hole mass and Eddington ratio is fitted to the observations. This probability distribution is given by an assumed BHMF, ERDF and the selection function of the survey. We also corrected for evolution within our redshift range, transforming the distribution functions to $z=0$. This maximum likelihood method also ensures the consistency of the derived BHMF and ERDF with the AGN luminosity function. We were able to put tight constraints on both the active black hole mass function and the Eddington ratio distribution function.

The Eddington ratio distribution function is well described by a Schechter function with low $\er$ slope $\alpha_{\er} \approx -1.9$. The data are consistent with no decrease of the ERDF at low $\er$, within the constrained range. Using a log-normal distribution, we found a maximum at $\log \er = -1.8$, what can be taken as an upper limit for a potential turnover in the ERDF. 
Our results clearly show a wide distribution of Eddington ratios, in contrast to a single value or to a narrow log-normal distribution, which is based on the observed distribution, without accounting for the underlying selection effects. While we also observe a narrow log-normal distribution of Eddington ratios, this is in agreement with the constrained Schechter function or wide log-normal distribution for the Eddington ratio distribution function, when survey selection effects are properly accounted for, because low-$\er$ objects will be systematically missed in flux limited samples.

The active BHMF is well described by different analytic models. In general, it strongly decreases at the high mass end and follows a power law at the low mass end with slope of $\alpha \approx -2$.
A good fit to the data is achieved by a function similar to a Schechter function, but modified by an extra parameter that determines the steepness of the high mass decrease.
We found no evidence for a decrease of the BHMF toward low mass, as indicated by \citet{Greene:2007a} for $\mbh \lesssim 10^{6.5} M_\odot$. However, our sample is not very sensitive in this low mass range.

We compared our local active BHMF with the local quiescent BHMF from \citet{Marconi:2004}, determining the active fraction, or duty cycle, of local black holes. This active fraction is decreasing with increasing black hole mass, consistent with a power law with slope $\sim-0.86$. Thus, the most massive black holes in the present universe are less active than their lower mass companions. At the highest \mbh only $10^{-4}$ of all black holes are currently in an active stage, i.e. accreting above 0.01 of the Eddington rate. This supports the general picture of anti-hierarchical growth of black holes. This mass dependence of the active fraction indicates that our assumption of an uncorrelated BHMF and ERDF cannot be sustained up to low values of $\er$ and thus we caution to extrapolate the distribution functions into the low $\er$ regime. Investigating a mass dependence of the ERDF would especially require a wider luminosity coverage of the sample.

By comparing our low~$z$ BHMF with the BHMF of a higher $z$-bin, presented by \citet{Vestergaard:2008} and \citet{Vestergaard:2009}, we found an indication that the most massive black holes are currently in a less active stage than at earlier cosmic times, also in general agreement with anti-hierarchical  black hole growth.

Recently, \citet{Marconi:2008} proposed a modified method to estimate \mbh, taking into account the effect of radiation pressure. So far, it is still unknown if radiation pressure has an important effect on the BLR or not \citep[see e.g.][]{Netzer:2008}. If we take into account radiation pressure and apply their \mbh estimation formula to our sample, the major effect is an increase of \mbh especially for the low~\mbh objects. In total, the dispersion of the \mbh distribution decreases from 0.65~dex to 0.63~dex. In the BHMF the space density at median \mbh increases, while at high \mbh the space density slightly decreases. This would strengthen even further the evidence for anti-hierarchical black hole growth. On the other hand it would change our observed \mbh, and especially our $\er$, distributions, and thereby our constrained BHMF and the Eddington ratio distribution function.

Our work strengthens the scenario of anti-hierarchical  growth of black holes, also seen in other studies \citep{Merloni:2004,Heckman:2004,Greene:2007a,Shankar:2007,Vestergaard:2009}, at least at low redshift. The observation of 'cosmic downsizing' in the X-ray luminosity function \citep[e.g.][]{Ueda:2003,Hasinger:2005}, as well as in the optical, radio and IR luminosity function \citep[e.g.][]{Hunt:2004,Cirasuolo:2005,Matute:2006,Croom:2009}, i.e. the flattening of the faint end slope of the luminosity function towards higher redshift, is explained by the shift of the typical black hole mass of an active accreting black hole toward lower mass.

The presented local active black hole mass function and Eddington ratio distribution function serve as a local anchor for future studies of both distribution functions. These will provide further information on the cosmic history of growth and activity of supermassive black holes.

\begin{acknowledgements}
We thank Isabelle Gavignaud, Bernd Husemann and Natasha Maddox for helpful discussions.
We acknowledge support by the Deutsche Forschungsgemeinschaft under its priority programme SPP1177, grants Wi~1369/23-1 and Wi~1369/23-2. 
\end{acknowledgements}

\end{document}